\documentstyle[amstex,amssymb,11pt]{article}
\renewcommand{\baselinestretch}{1}

\input epsf

\begin{document}
%\sloppy
%\input{pref}
\newcount\nummer \nummer=0
\def\f#1{\global\advance\nummer by 1 \eqno{(\number\nummer)}
      \global\edef#1{(\number\nummer)}}

%begin macros

\newcommand{\pref}[1]{\ref{#1}}
\newcommand{\plabel}[1]{\label{#1}}
\newcommand{\prefeq}[1]{Gl.~(\ref{#1})}
\newcommand{\prefb}[1]{(\ref{#1})}
\newcommand{\prefapp}[1]{Appendix~\ref{#1}}
\newcommand{\plititem}[1]{\begin{zitat}{#1}\end{zitat}}
\newcommand{\pcite}[1]{\cite{#1}}
\newcommand{\plookup}[1]{\hoch{\ref{#1}}}

\newcommand{\N}{\mbox{{\rm I \hspace{-0.865em} N}}} % die natuerlichen Zahlen
\newcommand{\Z}{\mbox{$\Bbb Z$}}                    % die ganzen Zahlen
\newcommand{\R}{\mbox{{\rm I \hspace{-0.86em} R}}}  % die reellen Zahlen
                         
\let\oe=\o
\def\smovs{\hbox{$\{\hbox{sector-moves}\}$}}
\def\osmovs{\hbox{$\{\hbox{sector-moves}\}^\uparrow$}}
\def\oG{\hbox{${\cal G}_+^\uparrow$}}
\def\bpmovs{\hbox{$\{\hbox{bifurcation point moves}\}$}}
\def\Rboost{\hbox{${\R}^{\rm (boost)}$}}
\def\Zflip{\hbox{${{\Z}_2}^{\rm (flip)}$}}
\def\ZPT{\hbox{${{\Z}_2}^{\rm (PT)}$}}

\def\sckai{\overset{\scriptstyle k}s_i}
\def\sczai{\overset{\scriptstyle 0}s_i}
\def\sclai{\overset{\scriptstyle l}s_i}
\def\sclakinv{{\smash{\overset{\scriptstyle l}s}_k}^{-1}}
\def\sckak{\overset{\scriptstyle k}s_k}
\def\sciakinv{{\smash{\overset{\scriptstyle i}s}_k}^{-1}}
\def\sckakmo{\overset{\scriptstyle k}s_{k-1}}
\def\scoaf{\overset{\scriptstyle 1}s_4}
\def\scoat{\overset{\scriptstyle 1}s_2}
\def\scoaz{\overset{\scriptstyle 1}s_0}

\def\Di{\displaystyle}
\def\nn{\nonumber \\}
\def\re{(\ref }
\def\i{{\rm i}}
\let\a=\alpha \let\b=\beta \let\g=\gamma \let\d=\delta
\let\e=\varepsilon \let\ep=\epsilon \let\z=\zeta \let\h=\eta \let\th=\theta
\let\dh=\vartheta \let\k=\kappa \let\l=\lambda \let\m=\mu
\let\n=\nu \let\x=\xi \let\p=\pi \let\r=\rho \let\s=\sigma
\let\t=\tau \let\o=\omega \let\c=\chi \let\ps=\psi
\let\ph=\varphi \let\Ph=\phi \let\PH=\Phi \let\Ps=\Psi
\let\O=\Omega \let\S=\Sigma \let\P=\Pi \let\Th=\Theta
\let\L=\Lambda \let\G=\Gamma \let\D=\Delta      

\def\wt{\widetilde}
\def\w{\wedge}
\def\0{\over } \def\1{\vec } \def\2{{1\over2}} \def\4{{1\over4}}
\def\5{\bar } \def\6{\partial }
\def\7#1{{#1}\llap{/}}
\def\8#1{{\textstyle{#1}}} \def\9#1{{\bf {#1}}}

\def\({\left(} \def\){\right)} \def\<{\langle } \def\>{\rangle }
\def\lb{\left\{} \def\rb{\right\}}
\let\lra=\leftrightarrow \let\LRA=\Leftrightarrow
\let\Ra=\Rightarrow \let\ra=\rightarrow
\def\ul{\underline}
                          
\let\ap=\approx \let\eq=\equiv
\let\ti=\tilde \let\bl=\biggl \let\br=\biggr
\let\bi=\choose \let\at=\atop \let\mat=\pmatrix
\def\CL{{\cal L}}\def\CX{{\cal X}}\def\CA{{\cal A}}
\def\CF{{\cal F}} \def\CD{{\cal D}} \def\rd{{\rm d}}
\def\rD{{\rm D}} \def\CH{{\cal H}} \def\CT{{\cal T}} \def\CM{{\cal M}}
\def\CI{{\cal I}} \newcommand{\dR}{\mbox{{\rm I \hspace{-0.86em} R}}}
  \newcommand{\dN}{\mbox{{\rm I \hspace{-0.865em} N}}}
\def\CP{{\cal P}}\def\CS{{\cal S}}\def\C{{\cal C}}

%%%%%% labels for references to parts I & II %%%%%%%%

\def\nullextr{{\bf II},20--22}
\def\eomRTa{{\bf I},30}
\def\OIIhchart{{\bf II},3}
\def\CasXsquX3{{\bf I},33,43}
\def\Kruskalref{{\bf II},33}
\def\saddleextr{{\bf II},29}
\def\saddleextrlength{{\bf II},30}
\def\flipref{{\bf II},17}
\def\JTref{{\bf II},11}
\def\KVref{{\bf II},13}
\def\JTetcref{{\bf II},11--13}
\def\deSitterref{{\bf II},10}
\def\Modelref{{\bf II},4}
\def\CcritKV{{\bf II},37}

%%%%%%%%%%%%%%%%%%%%%%%%%%%%%%%%%%%%%%%%%%%%%%%%%%%%%%%%%%%%%%%%%%%%%%

%end macros

\begin{titlepage}
\renewcommand{\thefootnote}{\fnsymbol{footnote}}
\renewcommand{\baselinestretch}{1.3}
\hfill  TUW - 96 - 06\\
\medskip
\hfill  PITHA - 96/17\\
\medskip
\hfill  hep-th/9607226\\
\medskip

\begin{center}
{\LARGE {Classical and Quantum  Gravity in 1+1 Dimensions\\
Part III: Solutions of Arbitrary Topology}
\\ }
\medskip
\vfill
            
\renewcommand{\baselinestretch}{1}
{\large {THOMAS
KL\"OSCH\footnote{e-mail: kloesch@@tph.tuwien.ac.at} \\
\medskip
Institut f\"ur Theoretische Physik \\
Technische Universit\"at Wien\\
Wiedner Hauptstr. 8--10, A-1040 Vienna\\
Austria\\
\medskip
\medskip THOMAS STROBL\footnote{e-mail:
tstrobl@@physik.rwth-aachen.de} \\ \medskip%\medskip
%\medskip \medskip
Institut f\"ur Theoretische Physik \\
RWTH-Aachen\\
Sommerfeldstr. 26--28, D52056 Aachen\\
Germany\\}}
\end{center}

\setcounter{footnote}{0}
\renewcommand{\baselinestretch}{1}                          %!!!
            
\begin{abstract}
All global solutions of arbitrary topology of the  most general 1+1
dimensional dilaton gravity models are obtained. We show that for a generic
model there are globally smooth solutions on any non-compact 2-surface.
The solution space is parametrized explicitly and the
geometrical significance of continuous and discrete labels is
elucidated. As a corollary we gain insight into the (in general non-trivial)
topology of the reduced phase space. \newline
The classification covers basically all 2D metrics of Lorentzian signature
with a (local) Killing symmetry. 

\medskip

\noindent PACS numbers: 04.20.Gz 04.60.Kz

\end{abstract}

\vfill
\hfill {\em Class.\ Quantum Grav.} {\bf 14} (1997), 1689.\\
\vfill
\end{titlepage}

\renewcommand{\baselinestretch}{1}
\small\normalsize

\section{Motivation and first results}
\plabel{Intro}

Much of the interest in two-dimensional gravity models 
centers around their quantization. However, for any interpretation 
of quantum results and, even more, for a comparison and 
possibly an improvement of existing quantization schemes, 
a sound understanding of the corresponding classical theory  
is indispensable.  

Therefore, in this paper we pursue quite an ambitious goal:
Given any 2D gravity Lagrangian of the 
form \cite{Odintsov}% FOOTNOTE
\footnote{In \re{L}) $g$ is a metric with Lorentzian signature 
  and $\Phi$ a scalar field, the `dilaton field'. 
  $R$ denotes the Ricci scalar. $D,V,Z$ are arbitrary (smooth)
  functions which, for technical reasons, we restrict by 
  $D' \neq 0$ and either $Z \neq 0$ or $Z \equiv 0$.}
\begin{equation}
  L[g,\Phi] = \int_M d^2 x \sqrt{|\det g|} \left[D(\Phi) R - 
  V(\Phi) + Z(\Phi) g^{\m\n} \6_\m \Phi \6_\n \Phi \right] \, ,
 \plabel{L}
\end{equation} 
we want to classify all its global, diffeomorphism
inequivalent classical solutions. This shall be done without
any restriction on the topology of the spacetime $M$.

For some of the popular, but specific choices of the potentials
$D,V,Z$, such as those of ordinary (i.e.\ string inspired, `linear')
dilaton gravity, of deSitter gravity, or of spherically reduced
gravity, cf.\ \cite{I}, the possible topologies of the maximally
extended solutions turn out to be restricted considerably through the
field equations. In particular their first homotopy is either trivial
or (at most) ${\Z}$. (Allowing e.g.\ also for conical singularities,
cf.\ Sec.\ \ref{Kinks}, the fundamental group might become more involved.)

For any `sufficiently generic' (as specified below) smooth/analytic 
choice of $D,V,Z$, on the other hand, the field equations of  $L$ allow for 
maximally extended, globally smooth/\hspace{0cm}analytic %% Trennhilfe
solutions on all non-compact two-surfaces with an arbitrary number of
handles (genus) and holes ($\ge1$).% FOOTNOTE
\footnote{That
   there are no solutions on compact manifolds (except in the flat case,
   cf.\ Sec.\ \ref{Const})
 can be seen by inspection of the possible fundamental groups;
 however, it may in many cases also be deduced from the fact that the range
 of the field $\Phi$ in \re{L}) is not compact. Let us note in this context
 that according to \cite{Thm} % p. 293
 there are {\em no\/} compact two-manifolds
 without boundary (closed surfaces) that may be endowed with a metric of
 Lorentzian signature, except for the torus and the Klein bottle.}
This shall be one of the main results of the present paper.
These solutions are smooth and maximally
extended, more precisely, the boundaries are either
at an infinite distance (geodesically complete) or they correspond to true
singularities (of the curvature $R$ and/or the dilaton field $\Phi$).
We will call such solutions {\em global\/}, as there are other kinds of
inextendible solutions (cf.\ below and Sec.\ \ref{Kinks}).

The existence of solutions on such non-trivial spacetimes is a
qualitatively new challenge for any programme of quantizing a gravity
theory. Take, e.g., a Hamiltonian approach to quantization: In any
dimension $D+1$ of spacetime the Hamiltonian formulation necessarily
is restricted to topologies of the form $\S \times \R$ where $\S$ is
some (usually spacelike) $D$-manifold. In our two-dimensional setting
$\S$ may be %either 
$\R$ or $S^1$ only. Thus $\pi_1(M)$ can be ${\Z}$
at most.  According to our discussion above this is far from
exhaustive in most of the models (\ref{L}). Let us compare this to the
case of full four-dimensional Einstein gravity. Clearly, there the
space of solutions will include spacetimes of rather complicated
topologies.  Therefore, a restriction to topologies of the
form $M =\S \times \R$ seems hardly satisfactory in the 4D scenario as well. 
A path integral approach to quantum gravity, on the other hand, does
not place an a priori restriction on the topology of the base manifold
$M$. However, also in this approach for $M \neq \S \times \R$ the
definition of an integration measure is plagued by additional
ambiguities and problems. The class of models \re{L}) may serve as a
good laboratory to improve on that situation and to gain new insights
in such directions.

For spacetimes of topology $\S \times \R$, furthermore, we are
interested in an explicit comparison of the solution space of \re{L})
(space of all solutions to the field equations modulo diffeomorphisms)
with the reduced phase space (RPS) in a Hamiltonian formulation of the 
theory. In the simply connected case ($\S = \R$) we already classified
all global diffeomorphism inequivalent solutions in \cite{II}. The
solution space was found to be {\em one\/}-dimensional, parametrized
by a real number $C \in \R$. As the result of a symplectic reduction
must lead to an {\em even\/}-dimensional RPS, we may
conclude that in the case of an `open universe' ($\S = \R$) the
proper definition of a Hamiltonian system, describing the same physics
as \re{L}), is in need of some additional external input. This may
creep in implicitly, e.g., when defining boundary/fall-off conditions
for the canonical phase space fields or may be introduced by
restriction to particular foliations. Periodic boundary conditions,
on the other hand, lead to a Hamiltonian formulation which is perfectly
well-defined without any further input besides that of 
periodicity (with respect to some arbitrarily fixed coordinate
period).  Effectively they describe the case of a `closed universe'
$\S = S^1$ and we conclude that for cylindrical topologies of $M$
the solution space of \re{L}) must be even-dimensional. Indeed it will
turn out to be two-dimensional, a second parameter `conjugate to $C$'
arising from the `compactification' (in one coordinate), cf.\ \cite{IV}.

For generic theories \re{L}) the solution space for $M \sim
S^1 \times \R$, and thus the corresponding RPS, will have a
highly non-trivial topology. This is the second challenge which has to be
faced in any quantization scheme: One has to cope with this non-triviality
of the orbit space, as for sure the RPS of four-dimensional
gravity will be even more intricate.

\medskip

Let us now
sketch how to describe the solution space of \re{L}) for arbitrary
topologies of $M$.
Our starting point will be the
universal covering solutions, % of the respective model, 
which we determined already in previous papers of this series
\cite{I,II}, referred to as {\bf I} and {\bf II}, respectively, in the
following: In {\bf I} we showed that for any of the models \re{L}) locally
$g$ may always be brought into a generalized Eddington-Finkelstein form
(Eq.\ (\OIIhchart)):
\begin{equation}
  g=2dx^0dx^1+h(x^0)(dx^1)^2 \, , \plabel{011h}
\end{equation}
in which case
$\Phi$ is a function of $x^0$ only. For the explicit
form of these two functions $h$ and $\Phi$ we refer the reader to
{\bf I}.% FOOTNOTE
\footnote{To generalize $Z\equiv 1/2$ to an arbitrary $Z>0$ just replace
  $\rho$ in ({\bf I},11) by $\rho = \int^\Phi Z(u)/D'(u) \, du$.}
Here we mention only that, up to diffeomorphisms, they are
determined completely in terms of the `potentials' $D,V,Z$, except for
{\em one\/} integration constant $C \in \R$. As an example, for
$D(\Phi) \equiv \Phi$, $Z(\Phi) \equiv 0$ one obtains
$h=\int^{x^0} V(u) du + C$ and $\Phi=x^0$.% FOOTNOTE
\footnote{Actually, the models considered in {\bf I}, {\bf II} were even more
  general than those of \re{L}). All of the above holds for
  generalizations of \re{L}) with non-trivial torsion as well. Also the
  results have been extended to the case of a (generally
  dilaton-dependent) coupling of \re{L}) to Yang-Mills fields of an
  arbitrary gauge group. In the latter case there arise additional
  parameters labelling the universal covering solutions (cf.\ {\bf I},
  {\bf II}) and certainly the solution space for other topologies
  changes, too. The content of the present paper may be adapted easily
  to this more general case, but for simplicity we discuss only models
  in the absence of Yang-Mills fields.}
With
formula \re{011h}) at hand we can now be more precise about the class
of models which allows for all non-compact two-surfaces:
This happens whenever the one-parameter family of functions $h$ of the
respective model contains functions $h$ with three or more zeros.
For the above $h=\int^{x^0} V(u) du + C$, obviously, this is the case,
iff the potential $V$ changes its sign at least twice.
We do not attempt to formulate the analogous conditions on
the three potentials $D,V,Z$ in \re{L}), since in {\bf I}
an explicit
formula for the one-parameter family $h$ has been provided and in
terms of the latter the condition for non-trivial spacetime topologies
is simple enough. 

In {\bf II} we constructed the maximal, simply connected extension of the
local solution \re{011h}). We showed that its global causal structure
is determined completely by the (number and kind of) zeros and the
asymptotic behaviour of the respective function $h$.
We derived elementary rules allowing for a straightforward
construction of the corresponding Penrose diagrams.  As a result one
obtains a one-parameter family of universal covering solutions, where
the shape of the Penrose diagrams changes with $C$ only when 
the function $h$ changes number and/or degree of its zeros
(or its asymptotic behaviour).% FOOTNOTE
\footnote{Actually, this is only fully true for the schematic diagrams
  of {\bf II} (disregarding, e.g., the curved boundaries of {\bf R1}
  in Figs.\ {\bf II},11 and {\bf II},7). However, these changes are irrelevant
  in the present context, since they do not influence the topology.} 

As an example we choose a model with
torsion, the KV-model \cite{Kat}, the Lagrangian of which consists
of three terms: curvature squared, torsion squared, plus a
cosmological constant $\L$.  As this model is well adapted to
illustrate much of what has been said up to now, we want to use it in
the following to collect first impressions of what to expect when
analysing the general model.

\begin{figure}[t]
\begin{center}
\leavevmode
\epsfxsize 7.5 cm \epsfbox{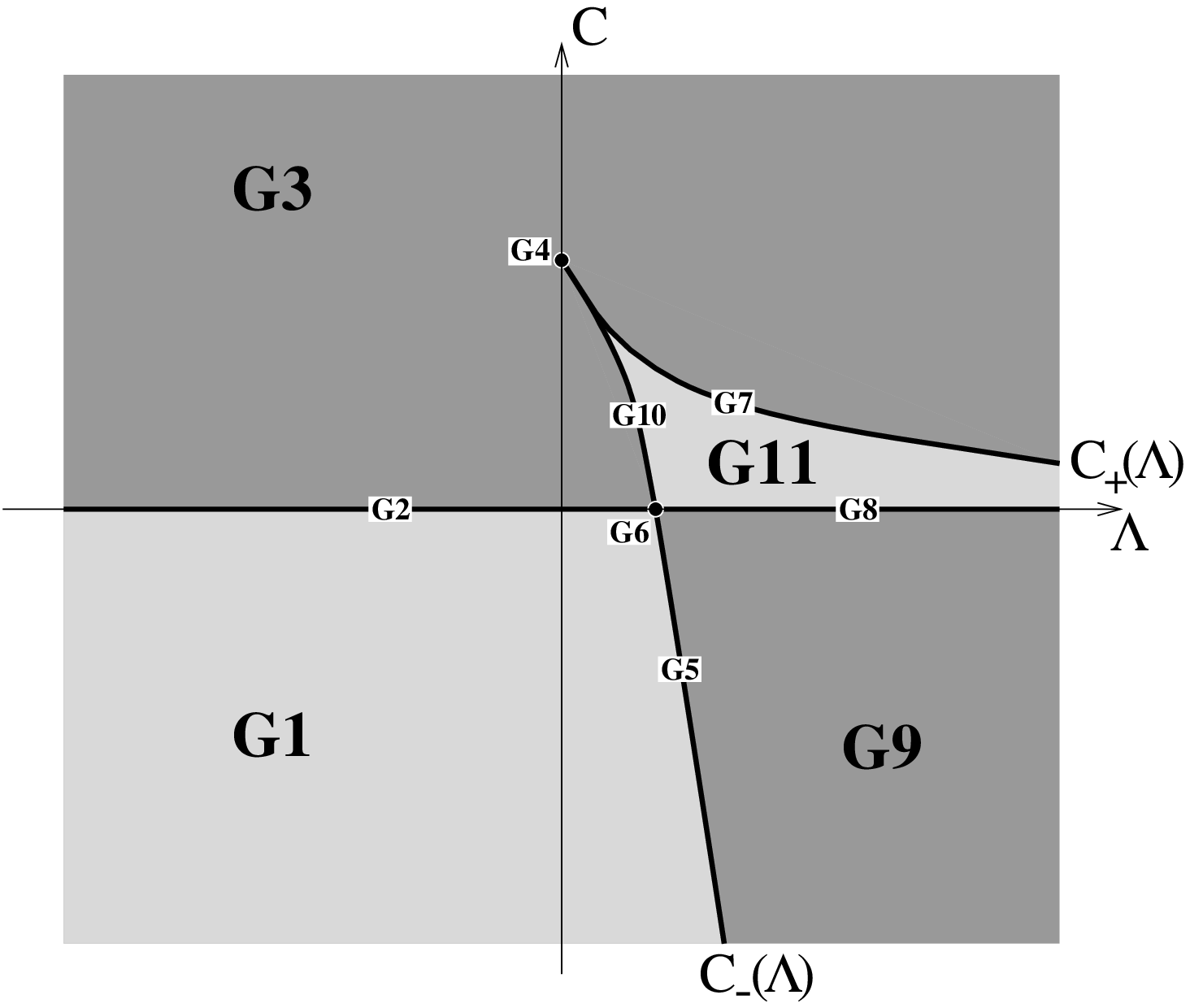}
\end{center}
\begin{quote}
{\bf Figure 1:} {\small KV-model, survey. The different regions
correspond to qualitatively different functions $h$ (number and degree of
zeros, asymptotic behaviour at 0 and $+\infty$). The Penrose diagrams for
{\bf G1,3,9,11} are given in Fig.\ 2.}
\end{quote}
\end{figure}

In the KV-model the function $h$ of \re{011h}) takes the form $h=C x^0
- 2 (x^0)^2 [(\ln x^0 - 1)^2 +1 - \L]$, Eq.\ ({\bf I},60), where $x^0 \in
\R^+$. Fig.\ 1\ shows a survey of the zeros of $h$ and its
asymptotic behaviour and thus a survey of the
various Penrose diagrams. For a negative cosmological constant $\L$
there are no zeros of $h$ for $x^0 \in \,  ]0,\infty[$ if $C<0$ and one
zero if $C>0$. The respective Penrose diagrams, {\bf G1} and {\bf G3},
are drawn in Fig.\ 2. Despite some differences the situation for
negative $\L$ reminds one of spherically symmetric vacuum gravity: Also
there a horizon is present only for positive Schwarzschild mass $m 
\sim C$. Moreover, like
the spherical model the KV-model with $\L<0$
belongs to the class of particular models where the possible
topologies of spacetimes are restricted severely.
As mentioned above one obtains all non-compact two-surfaces, if the
one-parameter family $h=h_C(x^0)$ contains functions with three or more
zeros. In the KV-model this is the case for positive $\L$. 
Correspondingly, there
are additional Penrose diagrams then: {\bf G11} if $\L>0$ as well as
{\bf G9} if $\L >1$, cf.\ Figs.\ 1,2.

\begin{figure}[t]
\begin{center}
\leavevmode
\epsfxsize 12 cm \epsfbox{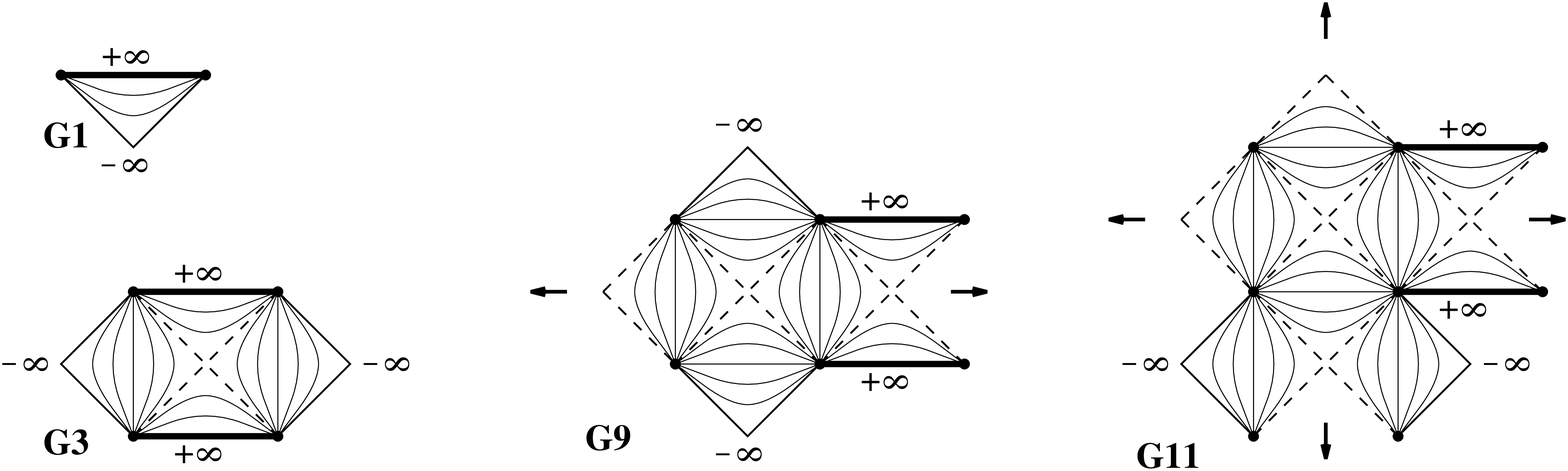}
\end{center}
\begin{quote}
{\bf Figure 2:} {\small Some Penrose-diagrams for the KV-model.
The thin lines denote Killing trajectories, the broken lines Killing
horizons. The style of the boundary lines indicates their completeness
properties; however, since they are irrelevant for our topological
considerations, we will treat them rather sloppily.
The arrows in {\bf G9} indicate that the patch should be extended horizontally
by appending similar copies, and likewise {\bf G11} sould be extended
vertically and horizontally.}
\end{quote}
\end{figure}

\medskip

Let us now discuss the possible global solutions that
correspond to the universal covering solutions in Fig.\ 2.
As will be seen later, the results depend only on the number and kind
of zeros of the function $h$. For reasons of brevity we skip the
solutions where $h$ has higher order zeros
in this introductory section; also we postpone a discussion of the deSitter
solutions (Sec.\ \ref{Const}). They both occur only for positive $\L$ and if
$C$ takes one of the two particular values $C_\pm(\L)$ on the boundary lines
between {\bf G1/G9} or {\bf G3/G11} (i.e.\ at {\bf G4-7,10} in Fig.\ 1).

We start with {\bf G1}:  Obviously this is a spatially homogeneous spacetime 
and \re{011h}) provides a global chart for it. Identifying 
$x^1$ and $x^1 + \o$, $\o=const$, evidently we obtain an 
everywhere smooth solution on a cylindrical spacetime 
$M \sim S^1 \times \R$. It results from 
the Penrose diagram {\bf G1} by cutting out a  (fundamental) region, e.g.\  
the strip between two null-lines in Fig.\ 3, and gluing both sides 
together in such a way that the values of the curvature scalar $R$, 
constant along the Killing lines, coincide at the identified
ends.  The constant $\o$ (or a function of it) becomes 
the variable conjugate to $C$ here. It may be characterized in
an inherently diffeomorphism invariant manner as the
(metric induced) distance between two identified points on a
line of an arbitrarily fixed value of $R$ (e.g.\ $R=0$). Thus $\o$
is a measure for the `size of the compact (spacelike) universe'.

\begin{figure}[t]
\begin{center}
\leavevmode
\epsfxsize 11 cm \epsfbox{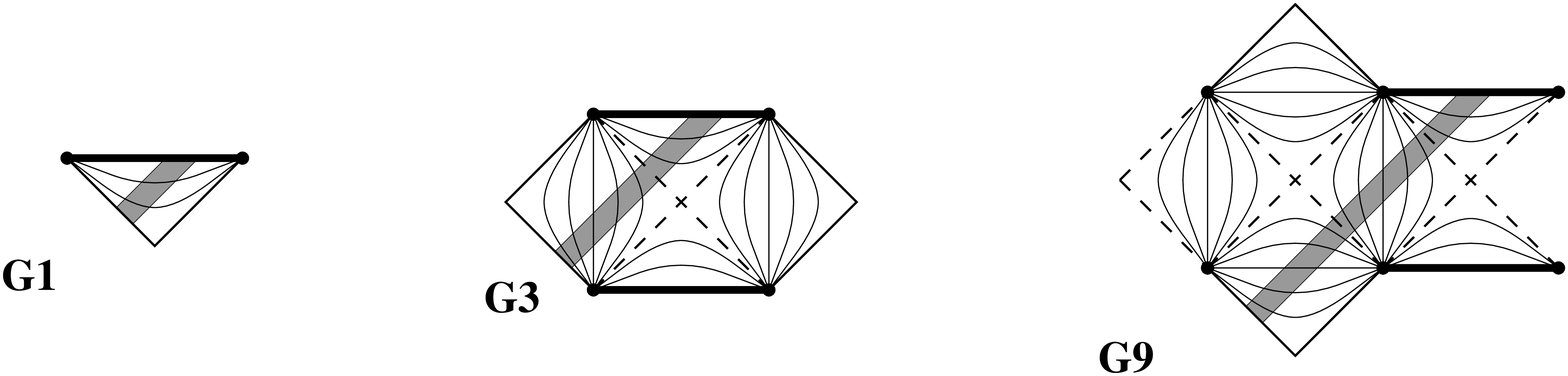}
\end{center}
\begin{quote}
{\bf Figure 3:} {\small Cylinders from {\bf G1,3,9}. The opposite
sides of the shaded strip have to be identified along the Killing lines
(cf.\ also Fig.\ 10). Note that in {\bf G3,9} there occur closed Killing
horizons (broken lines), which leads to pathologies of the Taub-NUT type.}
\end{quote}
\end{figure}

Next {\bf G3}: Clearly also in this case we can identify $x^1$ with $x^1+\o$ 
in a chart \re{011h}); obviously the resulting metric is completely smooth 
on the cylinder obtained, the fundamental region of which is drawn in 
Fig.\ 3.  However, this cylinder has some pronounced 
deficiencies:  
Not only does it contain closed timelike curves as well as one closed
null-line (the horizon); this spacetime, although smooth, is geodesically 
incomplete. There are, e.g., null-lines which wind around the cylinder
infinitely often, asymptotically approaching the horizon while having only
finite affine length (Taub-NUT spaces, see Sec.\ \ref{Factor}). So, from a
purely gravitational point of view such solutions would be excluded. Having
the quantum theory in mind, one might want to regard also such
solutions. Being perfectly smooth solutions on a cylinder, 
certainly they will  be contained in the RPS of the Hamiltonian theory. 
We leave it to the reader to exclude such solutions 
by hand or not.

The above Taub-NUT solutions are not the only dubious ones.
Take any maximally extended spacetime, remove a point from it, and
consider the $n$-fold covering of this manifold: clearly the
resulting spacetime is incomplete. However, for $n \neq 1$ it is
impossible to extend this manifold so as to regain the
original spacetime. This is a very trivial example of how
to obtain an in some sense maximally extended $2n$-kink solution
from any spacetime. In the presence of a Killing vector there
are, however, more intricate possibilities of constructing kinked
spacetimes (resulting, e.g., also in inextendible
2-kink solutions, even flat ones);
we postpone their discussion to Sec.\ \ref{Kinks}.% FOOTNOTE
\footnote{One justification for this separation (besides technical issues)
  is that the RPS of the
  Hamiltonian formulation introduced in \cite{IV} or \cite{Briefetc} is in
  some sense insensitive to these solutions (so effectively one may ignore
  them to find the
  RPS). This may be different in other Hamiltonian treatments. We will
  come back to this issue, cf.\ \cite{IV}.}
These solutions are certainly not global in the sense pointed out before.
Excluding them as well as the Taub-NUT type spaces from {\bf G3},
the above solutions are {\em all\/} global solutions for the KV-model
with negative cosmological constant $\L$. In particular we see that
the topology of spacetime is planar or cylindrical only.  Also
the RPS ($\S = S^1$) is found to have a simple structure: It is
a plane, parametrized by $C$ and $\o$.

Now the KV-model with $\L > 0$: The discussion of {\bf G1,3} is as
above. Also for {\bf G9,11} an identification $x^1 \sim x^1 + \o$ in a
chart \re{011h}) again leads to a smooth (but incomplete) cylinder.
However, for {\bf G9,11} there are also cylindrical solutions without
any deficiencies. Take, e.g., {\bf G9}: Instead of
extending the patch from Fig.\ 2\ infinitely by adding further copies
one could take only a finite number of them and glue the faces on the left and
the right % -hand ?
side together (cf.\ Fig.\ 4). Clearly the result will
be an everywhere smooth, inextendible
spacetime. Also it allows for a global foliation into
$\S \times {\R}$ with a spacelike $\S \sim S^1$. Now, however, 
the `size' of the closed universe is `quantized' (the circumference being
fully determined by the (integer) number of blocks involved).
Still, there is some further ambiguity in the gluing process that leads to a
one-parameter
family of diffeomorphism inequivalent cylinders for any fixed value of
$C$ and fixed block number. This second quantity, conjugate to
$C$ in a Hamiltonian formulation, and its geometric interpretation
shall be provided in the body of the paper (cf.\ Fig.\ 14\ below).  

\begin{figure}[t]
\begin{center}
\leavevmode
\epsfxsize 10 cm \epsfbox{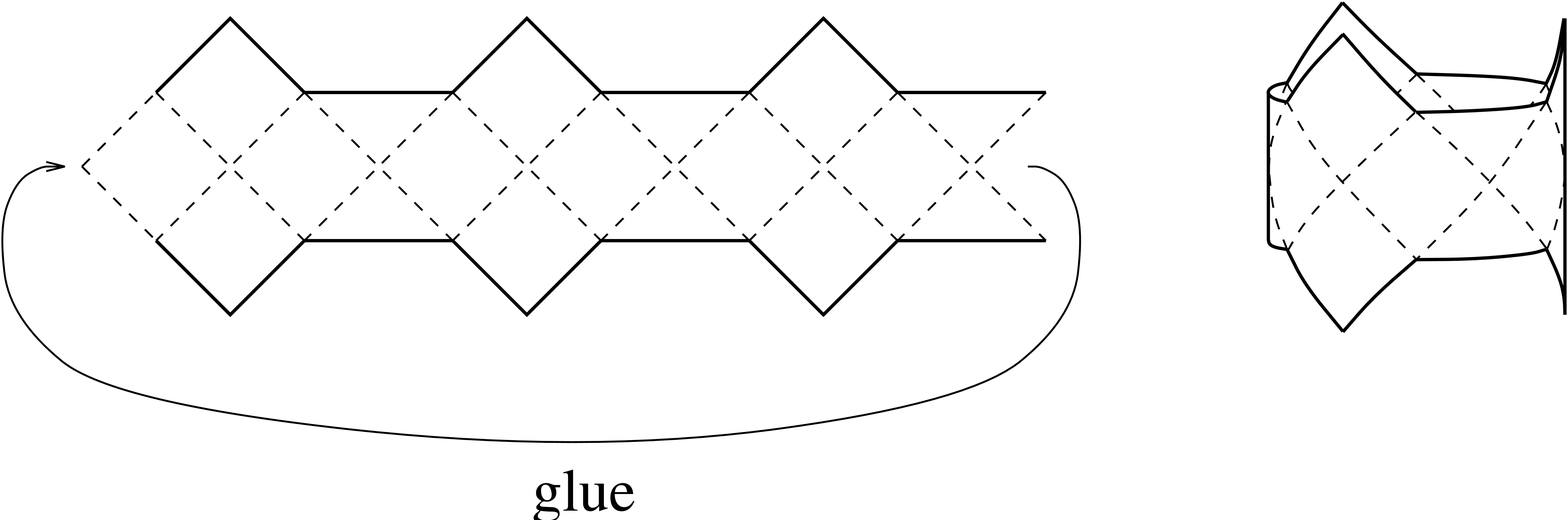}
\end{center}
\begin{quote}
{\bf Figure 4:} {\small Non-Taub-NUT cylinder from {\bf G9}.}
\end{quote}
\end{figure}

Thus in the case of {\bf G9} (and similarly of {\bf G11}) we find the
solution space for cylindrical spacetimes to be parametrized by $C$
(within the respective range, cf.\ Fig.\ 1), by an additional
real gluing parameter, {\em and\/} by a further discrete label (block
number).
For {\bf G9} there is also the possibility of solutions on a M\"obius
strip: We only have to twist the ends of the horizontal ribbon prior to
the identification. It will be shown that these non-orientable
solutions are determined uniquely already by fixing $C$ and the block
number; there is now no ambiguity in the gluing!

By far more possibilities arise for {\bf G11}. Again there are
cylinders of the above kind, with an analogous
parametrization of these solutions. However, now we can also identify
faces in vertical direction (cf.\ Fig.\ 2).
For instance, gluing together the upper and lower ends as
well as the right and left ends of the displayed region, one obtains a
global solution with the topology of a torus with hole. (It has
closed timelike curves, but no further defects; also there are tori
without CTCs).
The solution space for this topology is three-dimensional now, the two
continuous parameters besides $C$ resulting from inequivalent gluings
again. In addition, {\bf G11}
allows for much more complicated global solutions. In fact, it
is one of the examples for which solutions on all (reasonable)
non-compact topologies exist. Fig.\ 5\ displays two further
examples: A torus with three holes and a genus two surface with one
hole. The respective fundamental regions (for the topologically similar
solution {\bf R5}) are displayed in Fig.\ 
17\ below. As a general fact the dimension of the solution
space exceeds the rank of the respective fundamental group by one, and thus
it coincides with twice the genus plus the number of holes.  Also,
there occur further discrete labels.

\begin{figure}[t]
\begin{center}
\leavevmode
\epsfxsize 10 cm \epsfbox{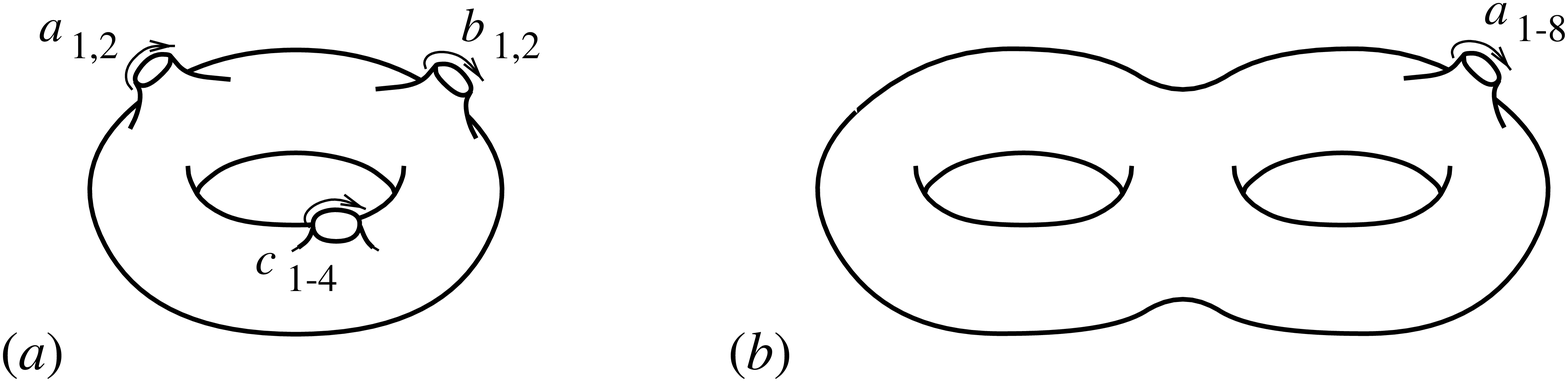}
\end{center}
\begin{quote}
{\bf Figure 5:} {\small Factor spaces from {\bf G11}. The
corresponding fundamental regions are similar to those given in
Fig.\ 17\ for {\bf R5}.}
\end{quote}
\end{figure}

In conclusion, let us consider the RPS (= the solution
space for topology $S^1 \times \R$) in the case of
$\L>1$ (simultaneous existence of {\bf G9}
and {\bf G11}, cf.\ Fig.\ 1).  Again it is two-dimensional,
being parametrized locally by $C$ and the respective conjugate
`gluing' variable. However, for $C$ taken from the open interval
$]C_-,C_+[$ (where $C_\pm(\L)= -4 \( \pm \sqrt\L -1 \) \exp\(\pm \sqrt{\L}\)$,
cf.\ Fig.\ 1\ and \CcritKV) there are infinitely many such
two-dimensional parts of the RPS, labelled by their
`block number'. More precisely, for sufficiently large negative
numbers of the (canonical) variable $C$ ($C < C_-$) the RPS consists
of one two-dimensional sheet. At $C=C_-$ this sheet splits into
infinitely many two-dimensional sheets. At $C=0$, furthermore,
any of these sheets splits again into infinitely many, all of
which are reunified finally into just one sheet for $C \ge C_+$.
Furthermore, at $C_\pm$ and $0$ the RPS is non-Hausdorff. 

So, while  for $\L <0$ a RPS quantization is straightforward, yielding 
wavefunctions $\Psi(C)$ with the standard inner product, an RPS quantization 
is not even well-defined for $\L>0$ (due to the topological
deficiencies of the RPS). In a Dirac approach to quantization
\cite{IV,Briefetc}, on the other hand, related problems are
encountered when coming to the issue of an inner product between
the physical wave functionals: For $\L>1$, e.g.,  the states are
found to depend on $C$, again, but for $C \in \;  ]C_-,0[ \,$
there is one further discrete label,
and for $C \in \; ]0,C_+[ \,$ there are even further labels. When no
discrete indices occur, as is the case for
$\L < 0$, an inner product may be defined by
requiring that the Dirac observable $\o$ conjugate to $C$ becomes
a hermitean operator when acting on physical states
\cite{Ashbuch}; this again leads to the Lebesgue measure $dC$
then. Such a simple strategy seems to fail for $\L>0$ (and also any generic 
case of \re{L})).
This is one of the points where an improvement of quantization schemes may
set in. 

\medskip

In our above treatment of the KV-model we used a
`cut-and-paste' technique to construct the maximally extended
solutions from the Penrose diagrams: we cut out some fundamental
region from the universal covering solutions and glued it together
appropriately.  In our classification for the general model
\re{L}) a more systematic, group theoretical approach shall be
applied. For that purpose we will determine the full isometry group
${\cal G}$ of the universal covering solutions first. This will be
achieved in Section 3, after collecting some basics in Section \ref{Prelim}.
As pointed out there, all global solutions
may be obtained as factor solutions of the
universal coverings by appropriate discrete subgroups of ${\cal G}$;
this analysis is carried out in Sections \ref{Factor},\ref{Two}. However,
the simple machinery does not quite apply for those solutions with three
Killing vectors, which describe spaces of constant curvature and are thus
discussed separately in Section \ref{Const}.
Section \ref{Kinks}, finally, treats the non-global inextendible solutions
described before.

\section{Preliminaries}
\plabel{Prelim}

The method employed for finding the multiply connected solutions will be to
factor the universal covering solutions
by a properly acting transformation group.  Let ${\cal G}:=
\mbox{\it Sym}({\cal M})$ be the symmetry group of the manifold
${\cal M}$. For any subgroup ${\cal H}\le{\cal G}$ we can construct the
factor- (or orbit-)space ${\cal M}/{\cal H}$ which consists of the
orbits ${\cal H}x$ ($x\in{\cal M}$) endowed with the quotient topology
(e.g.\ \cite{Massey}).
To pass from this approach to a cut-and-paste description
choose a fundamental region, i.e.\ a subset
${\cal F}\subseteq{\cal M}$ such that each orbit ${\cal H}x$ intersects
${\cal F}$ exactly once. The group action then dictates how the points of
the boundary of ${\cal F}$ have to be glued together (cf.\ Fig.\ 6).

\begin{figure}[htb]
\begin{center}
\leavevmode
\epsfxsize 11 cm \epsfbox{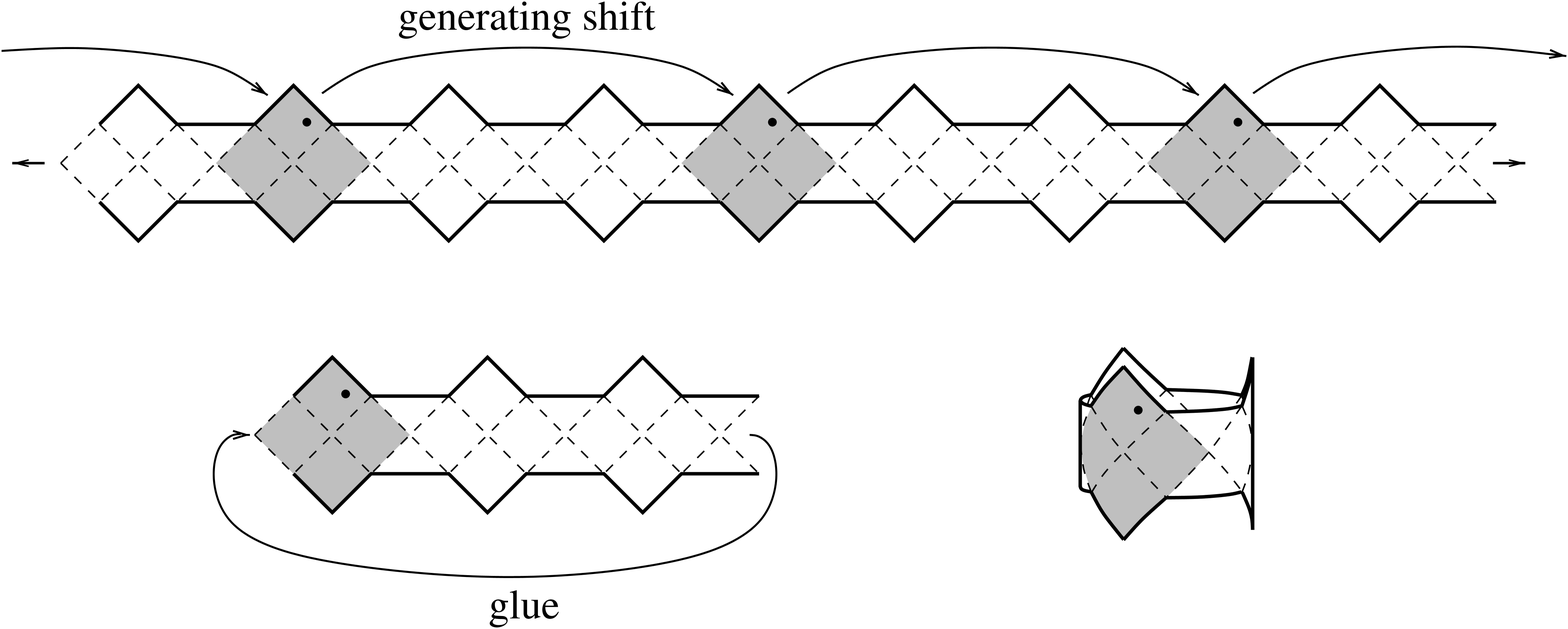}
\end{center}
\begin{quote}
{\bf Figure 6:} {\small Cut-and-paste approach versus factorization.
In the upper figure we indicated the action of a transformation group
generated by a shift three copies to the right. To obtain the orbit space
one identifies all sectors which are a multiple of three copies apart
(e.g.\ all shaded patches). This space may be described equivalently by
cutting out a fundamental region (lower figure) and gluing together the
corresponding faces.}
\end{quote}
\end{figure}

A priori orbit spaces may be topologically rather unpleasant,
they need e.g.\ not even be Hausdorff.
However, iff the action of this subgroup ${\cal H}$ is
free and properly discontinuous,% FOOTNOTE
\footnote{`Free action' in this context means fixed-point-free (not
  to be confused with `free group', which means that there are no
  relations between the generators of the group).  For the definition
  of properly discontinuous see e.g.\ \cite{Haw,Wolf,Massey}.  These
  two conditions on the action of $\cal H$ certainly imply that $\cal
  H$ is discrete with respect to any reasonable topology on $\cal G$.
  (The converse is, however, not true: a finite rotation group, e.g.,
  is discrete but has a fixed point).}
then the orbit space is again
locally ${\R}^n$ and Hausdorff, i.e.\ a manifold. If, furthermore,
${\cal H}$ preserves some (smooth, metric, etc.)  structure or fields
(e.g.\ $\Phi$), then the orbit space inherits such a structure in a
unique way, i.e.\ the metric and the other fields are well-defined on
${\cal M}/{\cal H}$ (they `factor through') and still fulfill the
equations of motion (e.o.m.).

In this way one can obtain new `factor'-solutions of the e.o.m. We
now want to sketch shortly that when starting in this way from the 
universal covering ${\cal M}$,  one obtains {\em all\/} multiply
connected global solutions:
Given a multiply connected manifold $M$, one can always 
construct the (unique) simply connected universal covering space
$\wt M$ and, furthermore, lift all the structure
(metric, fields) to it. Certainly, the lifted fields again satisfy the
e.o.m.\ (since these equations are purely local).
Also, the lifted geodesics are geodesics
on the covering space, and they have the same completeness properties.
Thus the universal covering
of a global solution is again a global solution of the e.o.m.\ and coincides
with ${\cal M}$ (which
was found to be determined uniquely, cf.\ {\bf II}). Conversely, the
original multiply connected solution $M$ can be recovered from
the universal covering $\wt{M}={\cal M}$ by factoring out the
group of deck-transformations.  Let us note in passing that the
fundamental group of a factor space is isomorphic to the group
factored out, $\p_1(M)\equiv\p_1({\cal M}/{\cal H})\cong{\cal H}$ 
(more on this in \cite{Massey}).

The solutions obtained by this approach are all smooth,
maximally extended, and Hausdorff. Of course, if one is less
demanding and admits e.g.\ boundaries (non-maximal extension),
conical singularities (failures of the differentiable
structure), or violation of the Hausdorff-property, then there
are many more solutions. We will shortly touch such
possibilities in Sec.\ \ref{Factor} (Taub-NUT spaces) and Sec.\
\ref{Kinks}.  On the other hand, 
from the point of view of classical general relativity even 
the globally smooth solutions may still have unpleasant
properties such as closed timelike curves or the lack of global
hyperbolicity (cf.\ the previous section). In any case, our strategy will be
to describe all of them;
if necessary they may be thrown away afterwards by hand.

\medskip

Let us shortly summarize what is needed from the first two papers 
{\bf I} and  {\bf II}
(while some knowledge of  {\bf II} may be useful, a reading of 
{\bf I} is not necessary): 
As remarked already in Section \ref{Intro} the solutions to the model 
\re{L}) or, more generally, 
to (\Modelref)  could be brought into the Eddington-Finkelstein
(EF) form \re{011h}) locally,  
with the dynamical fields $\Phi$ or $X^a,X^3$, respectively, depending
on $x^0$ only.  The zeros of $h$ denote Killing horizons and divide
the coordinate patch \re{011h}) into {\em sectors\/}. A Killing
horizon is called non-degenerate, if the corresponding zero of $h$ is
simple ($h'\ne0$), and degenerate otherwise. The patch was then
brought into conformal form, the `building block'
(Fig.\ 7\ ({\it b\/})). Let us note in this context that the shape
of the outermost sectors of this building block (square-shaped or triangular)
is irrelevant for our analysis.
\begin{figure}[t]
\begin{center}
\leavevmode 
\epsfxsize 14 cm \epsfbox{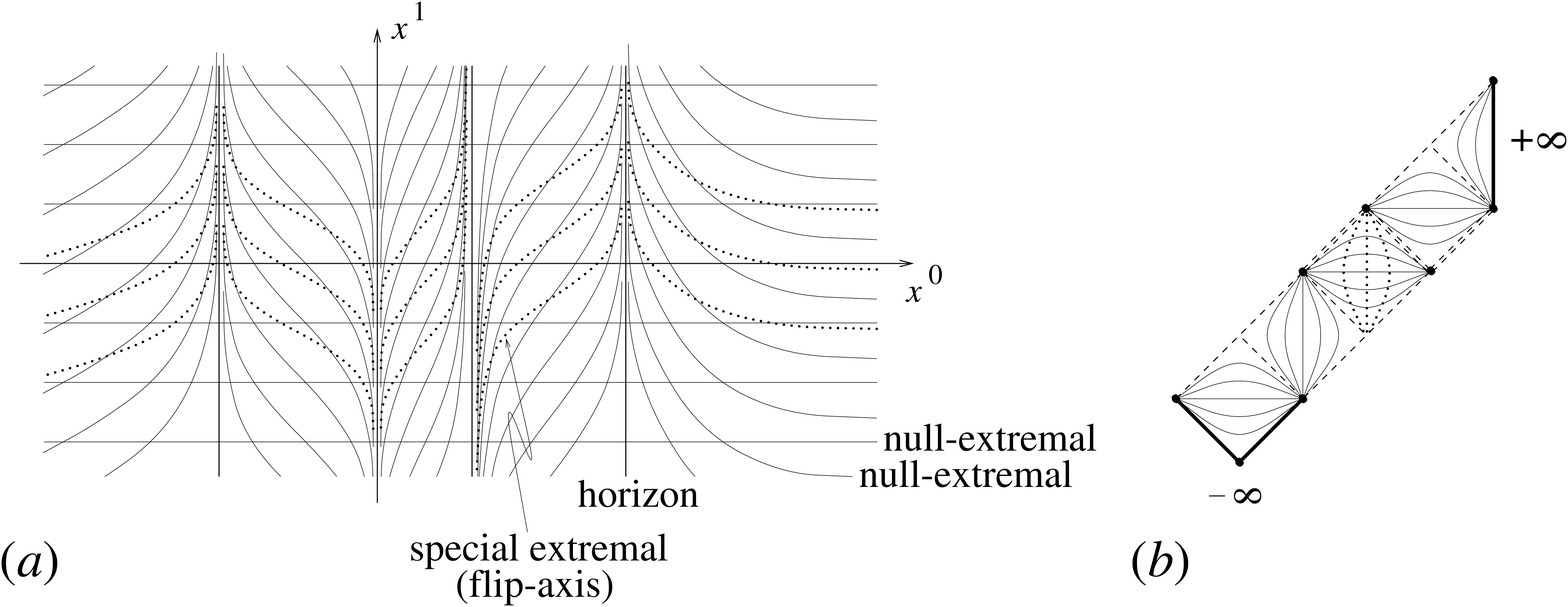}
\end{center}
\begin{quote}
 {\bf Figure 7:} {\small Building block. ({\it a\/}) EF-coordinates
 used in Eq.\ \re{011h}), ({\it b\/}) the corresponding (part of the) Penrose 
 diagram. In ({\it a\/}) the null-extremals \re{null1}--\ref{null2b}) have been
 drawn, which would run under $\pm 45^\circ$ in ({\it b\/}). The thin solid
 lines in ({\it b\/}) are the Killing trajectories, $x^0=const$ (vertical lines
 in ({\it a\/})). The broken lines in ({\it b\/}) are Killing horizons
 (multiple broken ones for degenerate horizons). Finally, the special extremals
 \re{special}) have been drawn in ({\it a\/}) and in one sector of ({\it b\/})
 as dotted lines.}
\end{quote}
\end{figure}

The metric \re{011h}) displays two symmetries, namely the
Killing field $\6\0\6x^1$, generating the transformations
\begin{equation}
  \tilde x^0 =x^0 \quad , \qquad \tilde x^1= x^1 + \omega\,,
  \plabel{Kill}
\end{equation}
valid within one building block, and the (local) {\em flip\/}
transformation
\begin{equation}
  \tilde x^0 =x^0 \quad , \qquad \tilde x^1= -x^1-2 \int^{x^0}
  \!\! {du\0h(u)} + \mbox{const}\,,
  \plabel{glue}
\end{equation}
valid within a sector (cf.\ \flipref), which has been used as gluing
diffeomorphism for the maximal extension.  The following extremals will be
of some interest: the null-extremals (cf.~\nullextr)
\begin{eqnarray}
  x^1&=&\hbox{const} \, , \plabel{null1} \\ {dx^1\over
    dx^0}&=&-{2\over h} \quad,\quad \mbox{wherever }h(x^0) \neq 0,
  \plabel{null2a} \\ x^0&=&\hbox{const} \, ,\quad \mbox{if} \,\,
  h(x^0)=0 \, , \plabel{null2b}
\end{eqnarray}   
and the special family of non-null extremals (cf.\ \saddleextr)
\begin{equation}
  {dx^1\over dx^0}=-{1\over h}, \plabel{special}
\end{equation}
which are also the possible symmetry axes for the flip-transformations
\re{glue}). % (dotted lines in Fig.\ 7).

The building block is usually incomplete (unless
there is only one sector) and has thus to be extended. 
This process (described in detail
in {\bf II}) consisted of taking at each sector the mirror image of the
block and pasting the corresponding sectors together (using the gluing
diffeomorphism \re{glue})).
Usually, overlapping sectors should not be identified, giving rise to
a multi-layered structure, cf.\ e.g.\ the spiral-staircase appearance
of {\bf G4} (Fig.\ 11\ below). Only where non-degenerate horizons
meet in the manner of {\bf G3} (Figs.\ 2,11, called
{\em bifurcate\/} Killing horizons), the enclosed vertex point is an interior
point (saddle-point for $\Phi$, called {\em bifurcation point\/}),
and the overlapping sectors have to be glued together, yielding one sheet.

\medskip

Any symmetry-transformation of a solution to the model \re{L})
must of course preserve the function $\Phi$ (more generally, for the model
(\Modelref) the functions $X^3$, $X^a$) and also scalar curvature
(and, if non-trivial, also torsion).
However, by means of the e.o.m.\ of \re{L}) the
scalar curvature may be expressed in terms of $\Phi$. Similarly, for
(\Modelref) curvature and torsion can by the e.o.m.\ (\eomRTa) be
expressed in terms of the functions $X^3$ and $X^a$. Moreover, since $X^a$
carries a Lorentz index, one only has to preserve $(X)^2
\equiv X^a X_a$, which in turn may be expressed in terms of $X^3$ via
another field equation (Casimir function $C\left[(X)^2,X^3\right]=
const$, cf.\ \CasXsquX3)).  Hence, in order  to preserve all the
functions above, it is sufficient to preserve  $X^3$ 
only.% FOOTNOTE
\footnote{One could also consider neglecting preservation of $\Phi$
  resp.\ $X^i$ and regard isometries only, e.g.\ when one is interested merely
  in a classification of all global $1+1$ metrics with one (local) Killing
  field. In cases where $R(X^3)$ is not
  one-to-one this may lead to further discrete symmetries. We will not
  discuss the additional factor spaces that can arise as a
  consequence.} 
[Recall that, in its specialization to vanishing
torsion,  (\Modelref) describes \re{L}) upon the identification 
$\Phi=D^{-1}(X^3)$;
so $\Phi$ is preserved, iff $X^3$ is in this case (as common throughout the
literature, $D$ is assumed to be a diffeomorphism). Thus, also for
notational simplicity, we  shall speak of $X^3$ only; 
readers interested merely in \re{L}) may, however, well replace `$X^3$'
by `$\Phi$' in everything that follows.]

\medskip

Finally, we shortly summarize some facts concerning free groups
(details can be found in \cite{Massey, CombGr}). A free group is a
group generated by a number of elements $g_i$ among which there are no
relations. The elements of the group are the words
${g_{i_1}}^{k_1}\ldots{g_{i_l}}^{k_l}$, subject to the relations
(necessitated by the group axioms) $g_i{g_i}^{-1}=1$ and
$g_i1=1g_i=g_i$ (the unit element 1 is the empty word). A word is
called {\em reduced\/}, if these relations have been applied in order to
shorten it wherever possible. Multiplication of group elements is
performed simply by concatenating
the corresponding words and reducing if necessary. While for a given
free group there is no unique choice of the free generators, their
number is fixed and is called the {\em rank\/} of the group.  Free
groups are {\em not\/} abelian, except for the one-generator group; if
the commutation relations $ab=ba$ are added, then one speaks of a
{\em free abelian\/} group.

Subgroups of free groups are again free. However, quite contrary to
what is known from free abelian groups and vector spaces, the
rank of a subgroup of a free group may be larger than that of the
original group.
The number of the cosets (elements of ${\cal G}/{\cal H}$) 
of a subgroup ${\cal H}\le {\cal G}$ is called
the {\em index\/} of ${\cal H}$ in ${\cal G}$. If this index is finite,
then there is a formula for the rank of the subgroup $\cal H$:
\begin{equation}
  \mbox{index $\cal H$}={\mbox{rank $\cal H$}-1 \0 \mbox{rank ${\cal G}$}-1}\,,
  \plabel{index}
\end{equation}
(cf.\ \cite{Massey,CombGr}). In particular, subgroups of finite index
have {\em never\/} a smaller rank than the original group.  On the
other hand, $\mbox{rank $\cal H$}-1 = n \cdot (\mbox{rank ${\cal G}$}-1)$
for some $n$ does not guarantee that the index of the subgroup
$\cal H$ is finite;% FOOTNOTE
\footnote{For instance, there are a lot of proper subgroups $\cal H$
  with the same rank as ${\cal G}$ (e.g.\ those generated by powers of
  the original free generators). However, none of them can be of
  finite index: If they were, then according to \re{index}) they
  should have index 1; but this means that there is no coset besides
  $\cal H$, thus ${\cal H}={\cal G}$, contrary to the assumption.}
still, the question of whether a given subgroup has finite index is
decidable (cf.\ \cite{CombGr}), but the algorithm is rather
cumbersome.

\section{The symmetry group}
\plabel{Sym}

As pointed out in the previous section, any symmetry transformation must
preserve the function $X^3$; thus sectors must be mapped as a whole onto
corresponding ones (i.e.\ with the same range of $X^3$). Since $X^3(x^0)$ is
always monotonic (except for the deSitter solutions, which are
therefore discussed separately in Sec.\ \ref{Const}), this has also the nice
consequence that within one building-block a sector cannot be mapped onto
another one. So each
transformation gives rise to a certain permutation of the sectors and
we can thus split it into ({\it i\/}) a sector-permutation and ({\it ii\/})
an isometry of a sector onto itself.  Furthermore, it is evident that the
whole transformation is already fully determined by the image of only
one sector (the transformation can then be extended to the other
sectors by applying the gluing diffeomorphism \re{glue})).

Let us start with ({\it ii\/}), i.e.\ determine all isometries of one sector
onto itself. Again $X^3$ must be preserved, so in the chart
(\ref{011h}) the map must preserve the lines $X^3=const
\Leftrightarrow x^0=const$. But also null-extremals must be mapped
onto null-extremals. This leaves two possibilities: If the
null-extremals \re{null1}) (i.e.\ $x^1=const$) are mapped onto
themselves, then the only possibility is an overall shift of the
$x^1$-coordinate, $x^1 \to x^1+\omega$, i.e.\ a Killing-transformation
\re{Kill}).  The gluing diffeomorphism (\ref{glue}) shows that such a
transformation extends uniquely onto the whole universal covering, and
that in all charts \re{011h}) it is represented as an $x^1$-shift of the same
amount (but on the `flipped' ones in the opposite direction!). 
In the neighbourhood of bifurcation points (simple zero of $h(x^0)$) we can
also use the
local Kruskal-coordinates (\Kruskalref), where the same transformation reads
$u \to u\exp(\frac{h'(a)}2\omega)$, 
$v \to v\exp(-\frac{h'(a)}2\omega)$, 
which looks in this case very much like a Lorentz-boost (cf.\ e.g.\ the
arrows around the bifurcation points in Figs.\ 14, 20\ 
below). We will (thus) call these Killing-transformations shortly {\em
  boosts\/} and denote them by $b_\o$.   
The composition law is clearly $b_\o b_\n=b_{\o+\n}$, so the boosts
form a group isomorphic to $\R$, which shall henceforth be denoted by
$\Rboost$.

If, on the other hand, the two families of null-extremals, \re{null1})
(i.e.\ $x^1=const$) and \re{null2a}) (i.e.\ $dx^1/dx^0=-2/h$), are
interchanged, then we obtain precisely the flip transformations. These
transformations are the gluing diffeomorphisms \re{glue}), but
this time considered as active transformations. Due to the constant in
\re{glue}) the flips come as a one-parameter family; however, they
differ only by a boost, i.e.\ given a fixed flip transformation $f$,
any other flip $f'$ can be obtained as $f'=f b_\o \equiv b_{-\o}f
\equiv {b_{\o/2}}^{-1}fb_{\o/2}$ for some $\o$.  We will thus consider
only {\em one\/} flip and denote it by $f$.  In the Penrose diagrams such
a flip is essentially a reflexion at some axis (horizontal or
vertical, according to sgn$\,h$; cf.\ Fig.\ 8) and it
is of course involutive (i.e.\ self-inverse, $f^2=1$).
Let us finally point out that while under a pure boost each sector is
mapped onto itself (the corresponding sector-permutation is the
identity), a flip transformation always (unless there is only one
sector) entails a non-trivial sector-permutation (which is clearly
self-inverse, since $f$ is).

\medskip

We now turn to task ({\it i\/}), the description of the
sector-permutations: As pointed out above, each transformation is
fully determined by its action on only one sector. Let us thus choose
such a `basis-sector'. If the transformation does not preserve
orientation,% FOOTNOTE
\footnote{By this we mean the orientation of the spacetime considered as
  a 2-manifold, not the {\em spatial\/} orientation.}
we may apply a flip at this basis-sector and we are left
with an orientation-preserving transformation.  For these, however,
the corresponding sector-permutation is already uniquely determined by
the image of the basis-sector, i.e., for each copy of the basis-sector
somewhere in the universal covering there is exactly one
sector-permutation moving the basis-sector onto that copy.  All these
orientation-preserving sector-permutations thus make up a discrete
combinatorial group (henceforth called \smovs), which  now shall be
described in more detail.

Choose also a `basis-building-block' and within this basis-block fix
the first (or better: zeroth) sector as basis-sector (we label the
sectors by $0,\ldots,n$ and the horizons between by $1,\ldots,n$).  By
a `basic move' across sector $i$ we mean the following (cf.\ Fig.\ 
8): Go from the basis-sector to the $i$th sector of the
basis-block and from there to the zeroth sector of the 
perpendicular (i.e.\ flipped) block.
The corresponding sector-move, mapping the basis-sector onto this
other sector, shall be denoted by $s_i$ (of course $s_0$ is the identity,
$s_0=1$).  Also the inverse basic moves can be easily described: One
has to perform the same procedure at the {\em flipped\/} basis-block
only (cf.\ Fig.\ 8).  An inverse basic move is thus the
conjugate of the original move by a flip, $fs_if={s_i}^{-1}$ (note
$f=f^{-1}$).  Here $f$ has been supposed to be a flip at the
basis-sector. Flips at other sectors can be obtained by composition
with sector-moves: $s_if=f{s_i}^{-1}$ is a flip at sector $i$ of the
basis-block (Fig.\ 8).

\begin{figure}[t]
\begin{center}
\leavevmode
\epsfxsize 14cm \epsfbox{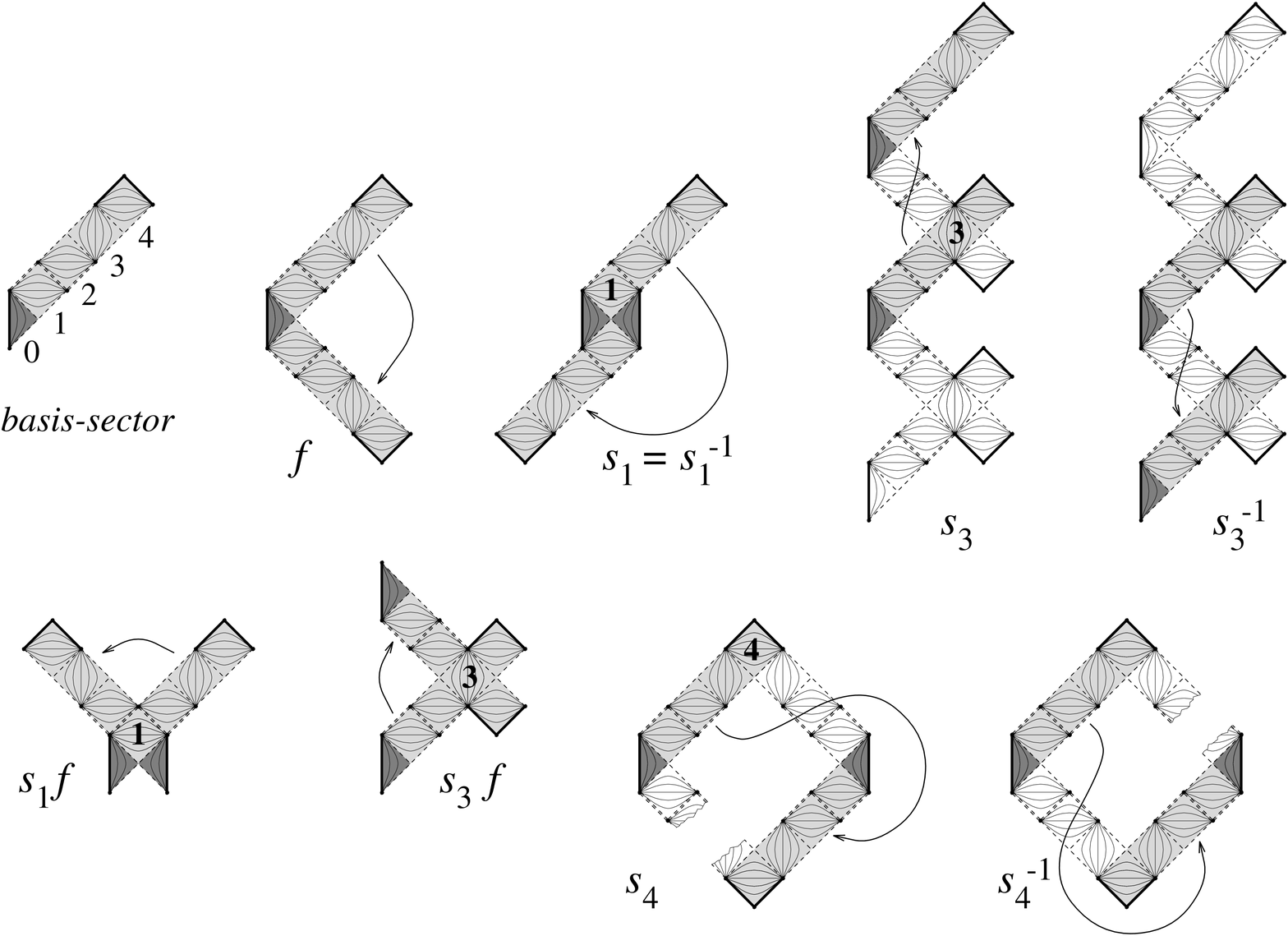}
\end{center}
\begin{quote}
  {\bf Figure 8:} {\small Some basic sector-moves, their
    inverses, and flips. The light shaded block is the basis-block
    resp.\ its image under the move, the dark shaded sector is the 
    basis-sector (or its image). Only a few sectors of the universal covering
    are displayed. Note that the basic move $s_4$ and its inverse
    ${s_4}^{-1}$ transport the basis-block into different (but
    overlapping) layers of the universal covering.
    And note also that $s_1$ (a bifurcation point reflexion), $s_4$,
    and their inverses turn the basis-block upside down!}
\end{quote}
\end{figure}

Evidently there are two qualitatively different cases: If basis-sector
and `flip'-sector are both stationary resp.\ spatially homogeneous, then the
basic move is essentially a translation (e.g.\ $s_3$ in Fig.\ 8).
However, if the sectors are of a different kind, then the move ($s_1$,
$s_2$, $s_4$ in the example of Fig.\ 8) turns the whole
solution upside down, inverting space {\em and\/} time (thus still
preserving the orientation, as required for elements of \smovs{} --- in
contrast to, e.g., $s_1 f$, which inverts space only, cf.\ Fig.\ 
8).

[Of course it is not necessary to choose the zeroth sector of the block as
basis-sector. Let us denote the basic move across sector $i$ with basis-sector
$k$ by $\sckai$ (thus $\sczai \equiv s_i$). 
They can, however, be expressed in terms of the old moves:
As may be seen from Fig.\ 9\ we have $\sckai=s_i{s_k}^{-1}$,
and consequently even $\sckai=\sclai\sclakinv$ for an arbitrary sector $l$.
Obviously always $\sckak=1$ (generalizing $s_0=1$) and $\sckai=\sciakinv$.
Thus, there is no loss of generality in choosing the basis-sector zero.]

\begin{figure}[htb]
\begin{center}
\leavevmode
\epsfxsize 10 cm \epsfbox{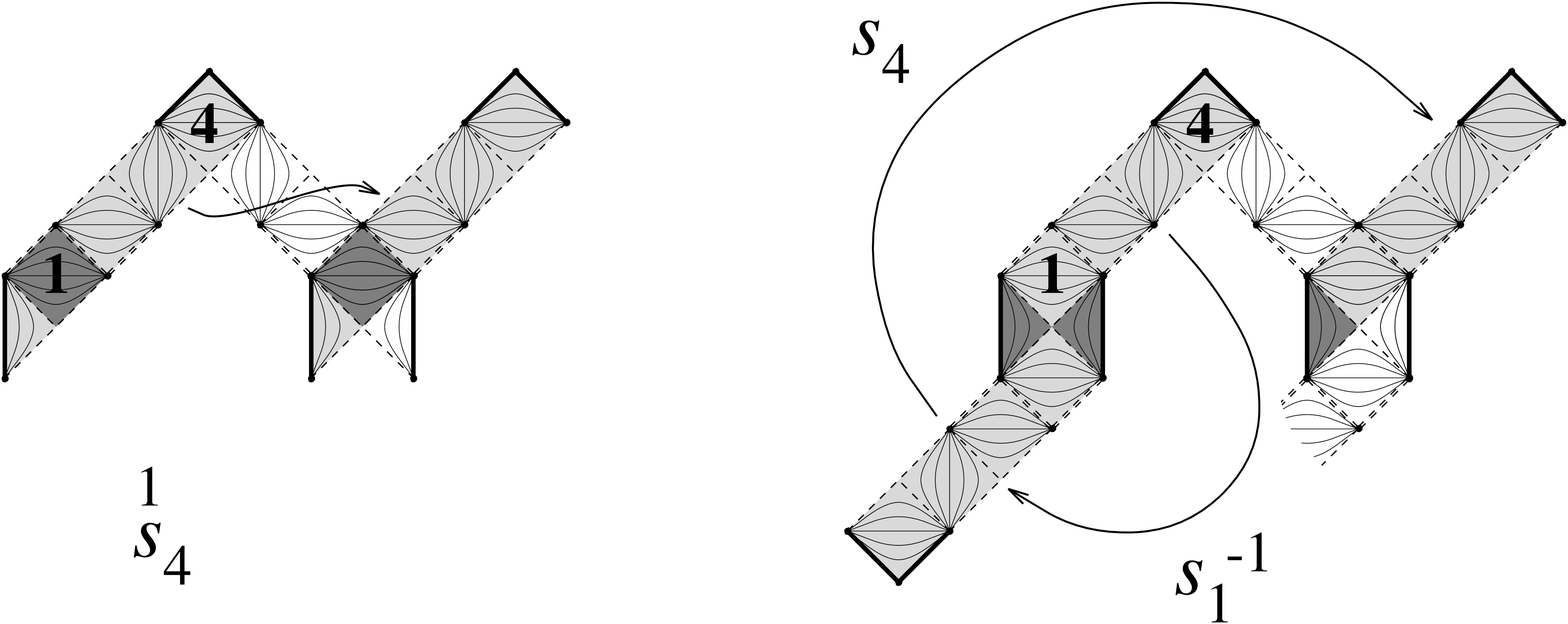}
\end{center}
\renewcommand{\baselinestretch}{.9}
\small \normalsize
\begin{quote}
{\bf Figure 9:} {\small Basic sector-move with different
basis-sector. As is seen from this figure the move $\scoaf$ can be composed
of two moves with basis-sector zero, $\scoaf=s_4{s_1}^{-1}$. Although in
this case ${s_1}^{-1}=s_1$ (relation at a bifurcation point), in general the
right element has to be an inverse move!}
\end{quote}
\end{figure}

The above basic moves $s_i$ form already a complete set of generators for
\smovs:
By the extension process each location in the universal covering can be
reached from the basis-sector by (repeated) application of the following
step: Move through the basis-block or the flipped basis-block to some sector,
then continue along the perpendicular block (i.e., applying a flip at this
sector). But this step {\em is\/} exactly a basic move.

There may yet be relations between the generators. If, however, all
horizons are degenerate, then there are no relations, and the
resulting group is the free group with generators $s_i$. The reason is
that in this case the vertex points between sectors are at an infinite
distance (cf.\ \saddleextrlength) and thus the overlapping sectors (after
surrounding such points) must not be identified, yielding a
multilayered structure (cf.\ e.g.\ $s_4$ in Fig.\ 8, or the
`winding staircase'-like {\bf G4} in Fig.\ 11\ below).  A
non-trivial word composed of basic moves $s_i$ describes such a move,
which must therefore necessarily lead into a different `layer' of the
universal covering, since this manifold is simply connected.

The situation is different if there are non-degenerate horizons: In this case
there are (interior!) bifurcation points at which four sectors meet, which
then constitute a single sheet. Moving once around such a bifurcation point
leads back to the original sector. Consequently there emerges a relation:
If e.g.\ the first horizon is non-degenerate then the basic move $s_1$ turns
the solution $180^\circ$ around with the bifurcation point as centre
(cf.\ Fig.\ 8). A second
application of $s_1$ yields the original configuration again, so we
have ${s_1}^2=1$. To find the analogous relations for a non-degenerate horizon
say between the $(k-1)$th and $k$th sector it is wise to
temporarily
shift the basis-sector to the $k$th sector. Then evidently $\sckakmo$ turns
the solution $180^\circ$ around that bifurcation point and we have the
relation $(\sckakmo)^2=1$ which
translates back to $(s_{k-1}{s_k}^{-1})^2=1$.

\medskip

Summarizing, we have found that the group \smovs\ has the following
presentation in terms of generators and relations ($n$ being the number of 
horizons):
\begin{equation}
\smovs = \left\<\; s_1,\ldots,s_n\;
     \Big|\; (s_{k-1}{s_k}^{-1})^2=1 
        \mbox{\ \ for each non-degenerate horizon $k$} \;\right\>
        \,\, .
\plabel{smovs}
\end{equation}

Any symmetry transformation can thus be written as a product of possibly
a flip $f$ (if it is orientation-reversing), a sector-move from the group
\re{smovs}), and a boost $b_\o$ ($\o\in{\R}$).
Furthermore, this representation is unique, provided the elements are in
this order.
We have yet to describe the group product:
Evidently boosts and sector-moves commute, since, as pointed out previously,
a boost is presented in all equally oriented charts (and sector-moves preserve
the orientation) as a shift $x^1\rightarrow x^1+\o$, Eq.\ \re{Kill}).
Furthermore, the conjugate
of a boost by a flip is the inverse boost, $fb_\o f=b_{-\o}={b_\o}^{-1}$,
and the conjugate of a basic move by a flip is the inverse basic move,
$fs_if={s_i}^{-1}$, which also 
defines the conjugate of a general (composite) sector-move.
The group product is thus completely determined, since 
in the formal product of two elements the factors can be 
interchanged to yield the above format.
Thus the structure of the full isometry group is a semi-direct product
\begin{equation}
  {\cal G}=\Zflip\ltimes \Big(\Rboost\times\smovs\Big),
  \plabel{symmgroup}
\end{equation}
where the right factor (in round brackets) is the normal subgroup,
$\Zflip$ denotes the group $\{1,f\}$,
and $\Rboost$ is the group of boosts described previously.
In particular, if we restrict ourselves to orientable factor spaces and
thus to orientation-preserving isometries, flips must be omitted and
we are left with a {\em direct\/} product of the combinatorial group
\smovs\ with $\Rboost$.

Still, this description is not always optimal, for two reasons: First,
one is often interested in orientable {\em and\/} time-orientable
solutions.  Second, while for only degenerate horizons the group
\smovs\ is a free group, this is not true if there are non-degenerate
horizons (cf.\ Eq.\ \re{smovs})).  Interestingly, both problems can be
dealt with simultaneously.  
Clearly the time-orientation-preserving sector-moves constitute a
subgroup of \smovs, which we shall denote by \osmovs\ (in analogy to the
notation used frequently for the orthochronous Lorentz group). If all
horizons are degenerate and of even degree, then all sectors are
equally `oriented' (stationary or spatially homogeneous) and consequently all
sector-moves are automatically time-orientation-preserving; thus
$\osmovs=\smovs$, and its rank equals the number of horizons.

This is no longer the case if there are horizons of odd degree.  Then
some sectors will have an orientation contrary to that of the
basis-sector and a sector-move at such a sector will reverse the
time-orientation (cf., e.g., $s_1$ in Fig.\ 8).  The group
\osmovs\ is then a proper subgroup of \smovs\ (with index 2), which
consists of all elements containing an even number of time-orientation
reversing sector-moves.

Fortunately, this group can be described quite explicitly.  Let us
start with the case that there are non-degenerate horizons.  To
simplify notation we assume for the moment the first horizon to be
non-degenerate (below we will drop this requirement).  The sector-move
$s_1$ is then an reflexion at a bifurcation point and thus inverts the
time-orientation (cf.\ Fig.\ 8).  But also the sector-moves
around all other sectors `oriented' contrary to the basis-sector
(sectors 2 and 4 in Fig.\ 8) will reverse the
time-orientation.  The idea is to extract the space-time-inversion
$s_1$ from the group \smovs: For each sector $i$ introduce the new
generators $s_i$ and $s_1s_is_1$, if the $i$th sector has the same
`orientation' as the basis-sector, and  $s_is_1$ and $s_1s_i$, if it
has the opposite `orientation'.  These new generators are all
time-orientation-preserving. Together with $s_1$ they obviously still span
the whole group \smovs\ (relations between the new generators will
be discussed presently). Conjugation by $s_1$ only permutes them among
themselves (use ${s_1}^2=1$): $s_is_1\leftrightarrow s_1s_i$ and
$s_i\leftrightarrow s_1s_is_1$.
Also, any element of \smovs\ can be expressed as a word composed of the new
generators with or without a leading $s_1$. Thus the group is a
semidirect product
\begin{equation}
  \smovs=\ZPT\ltimes\osmovs,
  \plabel{smovsplit}
\end{equation}
where $\ZPT=\left\{1,s_1\right\}\,$ (the `PT' stands for parity and
time-inversion) and \osmovs\ is the
group generated by the new elements given above (with $i \ge 2$).
There may still be some relations among these generators.
{}From \re{smovs}) we know that any non-degenerate horizon $k$ adds a relation
$(s_{k-1}{s_k}^{-1})^2=1$ or equivalently
$s_{k-1}{s_k}^{-1}=s_k{s_{k-1}}^{-1}$. This yields a relation between the
new generators, e.g.\ $(s_1s_{k-1}s_1)(s_ks_1)^{-1}=(s_1s_k){s_{k-1}}^{-1}$,
by means of which either of the (four) generators involved can be
expressed in terms of the remaining ones. 
Apart from those relations there are no
further ones, so if we eliminate the redundant generators we are left
with a {\em free\/} group!  To determine the rank of this group note
that each sector $>1$ contributes two generators and each
non-degenerate horizon (except for the first, which was taken into
account already in \ZPT) yields a relation which in turn kills one
generator. Thus
\begin{eqnarray}
\mbox{rank \osmovs} & = & 2 \;(\mbox{number of degenerate horizons})  +
                                                               \nonumber\\
            &   & \mbox{}+ \mbox{number of non-degenerate horizons} - 1 \,\,.
\plabel{rank}
\end{eqnarray}
Finally, if the first horizon is degenerate but the $k$th is
not, then one can replace $s_1$ above by $\sckakmo$ and proceed
in an analogous way. The result is \re{smovsplit},\ref{rank}) again.%
% FOOTNOTE
\footnote{Of course there may be other possibilities to split the isometry
  group into a product. For instance, if there are bifurcate horizons, then
  one can choose a basis-bifurcation-point and instead of tracking the motion
  of the basis-sector (which leads to \smovs) follow %(?)
  the bifurcation point. The resulting group of space- and
  time-orientation-preserving bifurcation point moves coincides exactly with
  \osmovs, which can thus also be interpreted as \bpmovs\
  (cf.\ e.g.\ Figs.\ 13,
  15 ({\it b\/}), 16). The remaining bifurcation point
  preserving isometries make up the 1+1 dimensional Lorentz group
  $O(1,1)$, which contains the boosts, flips (i.e.\ space-inversions and
  time-inversions), and reflexion at the bifurcation point
  (i.e.\ space-time-inversion). Hence the full isometry group can also be
  written as
  \[
   {\cal G}=O(1,1)\ltimes \bpmovs\,. 
  \]
  Furthermore, when restricting to 
  space- {\em and\/} time-orientable solutions we may use
  \[
    \oG=SO^\uparrow(1,1)\times\bpmovs
  \]
  where the proper orthochronous Lorentz group
  $SO^\uparrow(1,1) \cong \Rboost$ contains only the boosts. 
  Clearly, this is just the same as Eq.\ \re{osymmgroup}).}

The remaining case to consider is the one in which all the horizons are
degenerate. Then \smovs\ \re{smovs}) is a free group and correspondingly its
subgroup \osmovs\ is  free, too (cf.\ Sec.\ \ref{Prelim}). Let us determine 
its rank: As noted already above, if all the horizons are of
even degree, then $\osmovs\ = \smovs$, and the rank equals the total
number of (degenerate) horizons (thus \re{rank}) does not generalize to this
case). Finally, assume that there is an odd-degree horizon and let it again
be the first one (if it is not the first but the $k$th, just replace
$s_1$ by $\sckakmo$ in what follows): A set of generators can be
found in the same way as above (introducing $s_1s_k$, $s_ks_1$,
or $s_l$, $s_1s_ls_1$, respectively, $k,l
\ge 2$); however, now the element ${s_1}^2$, which no longer is the unit
element,  has to be added as a further generator. If $d$ denotes
the number of (degenerate) horizons, we thus find $2d -1$
generators for \osmovs\ (two for each degenerate horizon except
for the first horizon, which adds one only).  Here we also could
have used formula \re{index}), since
\osmovs\ has index 2 in \smovs.% FOOTNOTE
\footnote{Note, however, that in this case \smovs\ does not split into a
  semi-direct product like in \re{smovsplit}), since now there is no subgroup
  \ZPT (\smovs\ is free!); it is only a non-trivial extension of
  ${\Z}_2$ with \osmovs.}

Let us finally summarize the above results:
\begin{quote}
 \noindent{\bf Theorem:}\quad The group of space- and
 time-orientation-preserving symmetry transformations \oG\ is a direct product
 of $\R$ with a free group,
 \begin{equation}
   \oG=\Rboost\times\osmovs.
   \plabel{osymmgroup}
 \end{equation}
 Let furthermore $n$ denote the number of non-degenerate horizons and $d$
 the number of degenerate horizons.
 Then the rank of the free group \osmovs\ is 
 \begin{eqnarray}
      d   & \qquad\mbox{if all horizons are of even degree,}\nonumber\\
   2d+n-1 & \mbox{in all other cases.} 
   \plabel{oGrank}
 \end{eqnarray}
\end{quote}
We will mainly work with this subgroup, but nevertheless discuss at a few
examples how (time-)\-orientation-violating transformations can be
taken into account.

\section{Subgroups and factor spaces}
\plabel{Factor}

We now come to the classification of all factor solutions. 
As pointed out in Section \ref{Prelim}, they are obtained by factoring out a
freely and properly discontinuously acting 
(from now on called shorthand {\em properly acting\/})
symmetry group from the
universal covering solutions. Thus we have first to find all
properly acting subgroups of the full
symmetry group. However, not all different subgroups give rise to
different (i.e.\ non-isometric) factor spaces. If, for instance, two
subgroups $\cal H$ and ${\cal H}'$ are conjugate (i.e., there is a symmetry
transformation
$h\in{\cal G}$ such that ${\cal H}'=h{\cal H} h^{-1}$) then the factor spaces
${\cal M}/{\cal H}$ and ${\cal M}/{\cal H'}$ are %clearly?
isometric. (Roughly speaking, this conjugation can be interpreted as a
coordinate change).
But also the converse is true: 

\begin{quote}
 \noindent{\bf Theorem:} \quad Two factor spaces are isometric iff the
 corresponding subgroups are conjugate.
\end{quote}
(for a proof cf.\ e.g.\ \cite{Wolf}, Lemma 2.5.6).\\
The possible factor spaces are thus in one-to-one correspondence with the
conjugacy classes of properly acting subgroups.
Still, apart from this abstract characterization it would be nice to have 
some `physical observable', capable of discerning between the different factor
spaces. This concerns mainly the boost-components of the subgroup elements,
since information about the sector-moves is in a rather obvious way
encoded in the global causal structure (number and arrangement of the
sectors and singularities) of the factor space. 
Indeed we will in general be able to find such observables. 
[Here `physical observable' means some quantity that remains unchanged under 
the group of diffeomorphisms (= gauge symmetry); so it will 
be a {\em geometrical invariant\/}, which not necessarily can
be `measured' also by a `physical observer', strolling along his
timelike worldline].

The above theorem is also valid for the case of a restricted symmetry group.
For instance, a spacetime is often supposed to have --- apart from its metric
structure --- also an orientation and/or a time-orientation. If the universal
covering $\cal M$ carries a $\mbox{(time-)}$orientation and $\cal H$
is (time-)\-orientation-preserving, then also the factor space inherits a
(time-)\-orientation. Now, if two subgroups are conjugate,
${\cal H}'=h{\cal H} h^{-1}$, but the intertwining transformation $h$
does not preserve the (time-)\-orientation, then the corresponding factor
spaces will
have different (time-)\-orientations (while still being isometric, of course),
and should thus be regarded as different. Thus, in this case the conjugacy
classes of subgroups should be taken with respect to the restricted symmetry
group (e.g.\ \oG).

\medskip

The requirement that the subgroup acts freely
already rules out some transformations. First of all the subgroup must not
contain pure flips at any sector: A flip has a whole line of fixed points,
namely an extremal of the kind $dx^1/dx^0=-1/h$ (cf.\ \re{special}) and
Fig.\ 7) as the
symmetry axis. This symmetry axis would become a boundary line of the 
factor space, which then would no longer be maximally extended.
Consequently not only $f$, but also $f{s_i}^n$ and their conjugates have
to be omitted.% FOOTNOTE
\footnote{As pointed out previously $s_if$ is a flip at sector
  $i$. By the group product one has further $f{s_i}^{2k}={s_i}^{-k}f{s_i}^k$
  and $f{s_i}^{2k-1}={s_i}^{-k}(s_if){s_i}^k$, so they are conjugate to pure
  flips and thus also pure flips at displaced sectors.}

But also reflexions at a bifurcation point (turning the solution $180^\circ$
around this bifurcation point, e.g.\ $s_1$ in Fig.\ 8) must
be avoided: Not only does the factor space fail to be
time-orientable in this case, but also the bifurcation point is a
fixed point, which upon factorization develops into a singular
`conical tip' making the solution non-smooth.  Thus, if the
$k$th horizon is non-degenerate, then the elements $\sckakmo=
s_{k-1}{s_k}^{-1}$ and conjugates thereof must not occur in the
subgroup $\cal H$.

\begin{figure}[htb]
\begin{center}
\leavevmode
\epsfxsize \textwidth \epsfbox{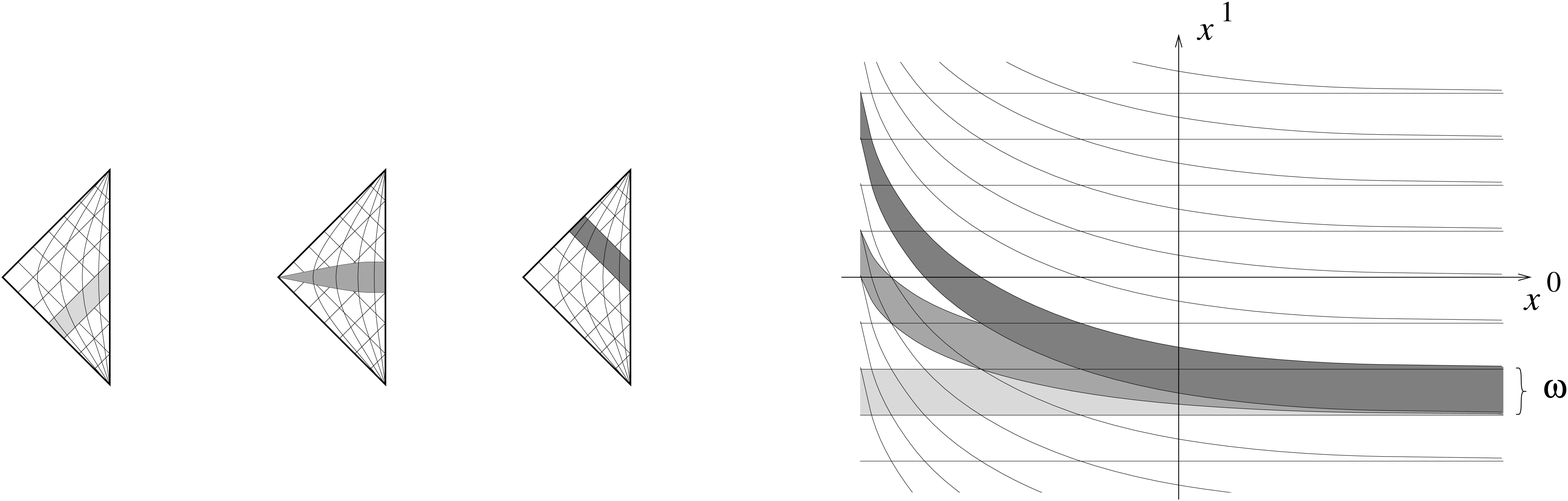}
\end{center}
\renewcommand{\baselinestretch}{.9}
\small \normalsize
\begin{quote}
{\bf Figure 10:} {\small A cylinder resulting from pure boosts and possible
fundamental regions $\cal F$; left-hand side in the Penrose diagrams,
right-hand side in the
EF-chart \re{011h}). Boundaries have to be glued along Killing-trajectories,
i.e.\ along the curved lines in the left, resp.\ along vertical fibres in the
right diagram.}
\end{quote}
\end{figure}

Let us next assume pure boosts, which form a group isomorphic to $\R$
(\Rboost).
The only discrete subgroups of boosts are the infinite cyclic groups
generated by one boost, ${\cal H}_\o := \left\{{b_\o}^n,\, n\in{\Z}
\right\},\: \o>0$.
In the coordinates (\ref{011h}) such a boost $b_\o$ is a shift of length $\o$
in $x^1$-direction. The factor space is then clearly a cylinder.
Also, since boosts commute with sector-moves and `anticommute' with flips
($fb_\o=b_{-\o}f$), the group ${\cal H}_\o$ is invariant under conjugation
and the parameter $\o$ cannot be changed. Thus the
cylinders are labelled by one positive real parameter $\o$. In order to find
some geometrical meaning of this parameter it is useful to adopt a
cut-and-paste approach to the factorization procedure 
(used already intuitively in Sec.\ \ref{Intro}): In the above example a
fundamental region $\cal F$ can be obtained from the patch
\re{011h}) by
cutting out a strip of width $\o$ parallel to the $x^0$-axis and gluing it
together along the frontiers (preserving $x^0$, i.e.\ vertical fibres).
Of course there are several choices for
${\cal F}$; it need not even be a horizontal strip, but any strip
with vertical cross-section $\o$ will work (cf.\ Fig.\ 10).
The width $\o$ of this strip (i.e.\ of the generating shift) is proportional to
the length of an $X^3=const$-path (resp.\ constant curvature or $\Phi$) 
running once around the cylinder.
This is the desired geometrical observable: for any metric $g$
(i.e.\ function $h$ in \re{011h})) we get a set of distinct solutions
parametrized by their `size' or circumference (any positive real 
number).% FOOTNOTE
\footnote{In Sec.\ \ref{Intro} we took the second continuous parameter 
  besides $C$ without restriction on its sign. For a comparison with the RPS 
  this is more appropriate, since $\o \sim -\o$  is induced 
  by the {\em large\/} diffeomorphism $x^1 \to - x^1$, which is not connected
  to the identity and thus cannot be generated by the flow of the constraints.}
Let us finally point out that this parameter $\o$ has nothing
to do with the Casimir-value $C$, present in the function $h$ of the metric.
It is a new, {\em additional\/} parameter resulting from the factorization.

This all works perfectly well as long as there is only one sector, i.e.,
$h$ has no zeros, but at horizons the boosts do not act properly
discontinuous. As a consequence the factor space will not be Hausdorff
there.
Furthermore, at bifurcation points the action is not free, so the factor
space is not even locally homeomorphic to $\R^2$ there.
At the first glance this might seem surprising since when starting from an
EF-coordinate patch \re{011h}) the construction of Fig.\ 10\ should yield
regular cylinders. They are not, however, global:
`half' of the extremals approach the (closed) horizon asymptotically,
winding around the cylinder infinitely often while having only
finite length.
This phenomenon is well-known from the Taub-NUT space or
its two-dimensional analogue as described by Misner
\cite{Mis} (cf.\ also \cite{Haw}).
A detailed description with illustrations can also be found in \cite{Geroch}.
[In {\bf G3}, Fig.\ 11, for instance, let the EF-coordinates cover
the sectors I and II. The above class of incomplete extremals then comprises
those which run across the lower horizon from I to IV, leaving
the EF-patch. Thus to obtain an extension a second half-cylinder
corresponding to IV has to be attached to I via a second copy of the closed
horizon and likewise for the sector III, but of course this violates the
Hausdorff property.]
Similar results hold in general for solutions with zeros of $h(x^0)$:
whenever the group factored out contains a pure boost we obtain a
Taub-NUT like solution (a cylinder-`bundle', where at each horizon two sheets
meet in a non-Hausdorff manner) labelled by its `size' (metric-induced
circumference along a closed $X^3=const$ line). If there are
further sector-moves in the group $\cal H$, they have only the effect of
identifying different sheets of this cylinder-bundle.

\medskip

After these preliminaries we will now start a systematic classification of
the factor solutions. We do this in order by number and type of horizons
and illustrate it by the solutions of the JT-, $R^2$-, and KV-model
(examples {\bf J1-3}, {\bf R1-5}, and {\bf G1-11}, respectively; for the
definition of the models and for all their Penrose diagrams cf.\ 
{\bf II}, but also Figs.\ 2--4,11--20).

\medskip
\noindent{\bf No horizons (e.g.\ G1,2, J3):}\\
\noindent This case has already been covered completely by the above
discussion: There is only one sector, so \smovs\ is trivial. Furthermore,
flips are not allowed, thus only boosts remain and they yield
cylinders labelled by their circumference (positive real number).

\medskip
 
In the following we will continue studying all factorizations possible
for solutions {\em with\/} horizons.
As pointed out before, there exist pathological Taub-NUT-like solutions for
all of these cases (resulting from pure boosts).
We will from now on exclude them, so
pure boosts must not occur in the subgroup $\cal H$. But this also means
that no sector-permutation can occur twice with different 
boost-parameters,
since otherwise they could be combined to a non-trivial pure boost.
This suggests the following strategy:
factor out $\Rboost$ from $\cal G$, i.e.\ project
the whole symmetry group onto ${\cal G}/\Rboost=\Zflip\ltimes\smovs$,
respectively \oG\ onto $\oG/\Rboost=\osmovs$. All above (non-Taub-NUT)
subgroups $\cal H$ are mapped one-to-one under this projection.
One can thus `forget' the
boost-components, and first solve the combinatorial problem of finding the
projected subgroups $\overline{\cal H}$ (which we will nevertheless simply
denote by $\cal H$).
Only afterwards we then deal with
re-providing the sector-moves with their boost-parameters.

\medskip
\noindent{\bf One horizon (e.g.\ G3-6, R1,2, J2):} (Fig.\ 11)\\
\noindent%
The group \smovs\ is in these cases still rather simple:
$\<s_1|{s_1}^2=1\>\cong{\Z}_2$ for a non-degenerate horizon
({\bf G3, R1}), and $\<s_1|-\>\cong{\Z}$ for the others.
First of all we see that flips have to be omitted altogether:
The most general form in which they can occur is $f{s_1}^kb_\o$, but this is
by the group product always
conjugate to $f$ or $s_1f$ and thus a pure flip.

\begin{figure}[htb]
\begin{center}
\leavevmode
\epsfxsize \textwidth \epsfbox{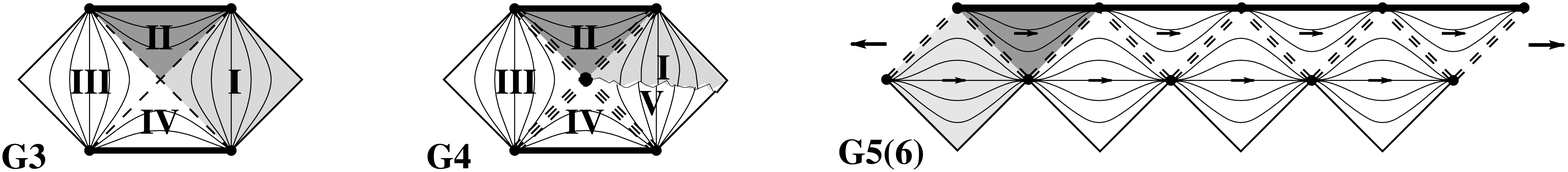}
\end{center}
\renewcommand{\baselinestretch}{.9}
\small \normalsize
\begin{quote}
{\bf Figure 11:} {\small Penrose diagrams for one horizon
({\bf G3-6}). For a non-degenerate horizon ({\bf G3}) the vertex point is an
(interior) bifurcation point and the manifold a single sheet. For degenerate
horizons of odd degree (e.g.\ {\bf G4}), on the other hand, this point is at
an infinite distance, and thus the manifold has infinitely many layers.
Finally, for horizons of even degree it is an infinite ribbon ({\bf G5(6)},
interior arrows indicate the Killing field).}
\end{quote}
\end{figure}

The action of \smovs\ on the manifold is most evident for a zero of even
degree, in which case we have the ribbon-structure of {\bf G5(6)}.
Then $s_1$ is a shift of one block to the right (in the situation of
Fig.\ 11). The non-trivial subgroups are the cyclic groups generated
by ${s_1}^n$, for some $n\ge 1$, and the corresponding factor space is obtained
by identifying blocks which are a multiple of $n$ blocks apart, i.e., gluing
the $n$th block onto the zeroth. This yields a cylinder of a `circumference'
of $n$ blocks (see Fig.\ 12\ for $n=3$). For degenerate horizons of
odd degree (e.g.\ {\bf G4}) the
situation is rather similar, only the basic move is now a
$180^\circ$-rotation around the (infinitely distant) central point,
`screwing the surface up or down' (i.e.\ mapping I $\rightarrow$ III,
II $\rightarrow$ IV, III $\rightarrow$ V, etc.). Thus we get again cylinders
of $n$ blocks circumference.
When passing once around such a cylinder, however, the lightcone tilts upside
down $n$ times, so we have got an {\it$n$-kink\/}-solution. In particular,
the solutions with $n$ odd
are not time-orientable. Finally, for non-deg.\ horizons (e.g.\ {\bf G3})
the only non-trivial sector-move is $s_1$ (${s_1}^2$ is already the
identity again), but this is the reflexion at the bifurcation point 
and thus not a free action. So we get no smooth factor solutions in this case.

If we had restricted ourselves a priori to orientable and time-orientable
solutions, then we would have had to start from \osmovs. For {\bf G5(6)}
this makes no difference since then $\osmovs=\smovs$, and indeed all cylinders
were time-orientable in this case. For {\bf G4}, on the other hand, \osmovs\
is a proper subgroup, $\osmovs\cong2{\Z}<{\Z}\cong\smovs$,
generated by ${s_1}^2$. This reflects the
fact that only solutions of even kink-number are time-orientable. Finally,
for {\bf G3} \osmovs\ is trivial and there are no factor solutions at all.

\begin{figure}[htb]
\begin{center}
\leavevmode
\epsfxsize 14cm \epsfbox{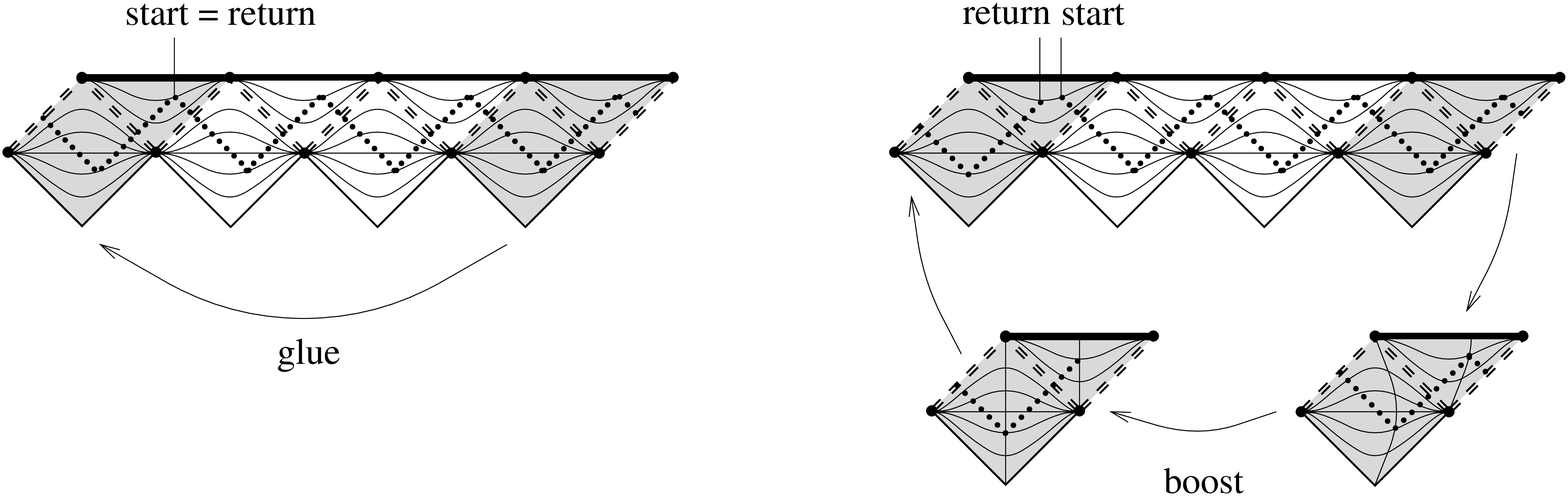}
\end{center}
\renewcommand{\baselinestretch}{.9}
\small \normalsize
\begin{quote}
{\bf Figure 12:} {\small Boost-parameter for {\bf G5(6)}. In both
cases the generating shift ${s_1}^3$ (or its inverse) is used to glue
the right-hand shaded patch onto the left-hand one. If this generator has a
non-trivial boost-component, then one has to apply a boost before gluing the
patches (right-hand figure).
To illustrate the effect we have drawn a polygon of null-lines (dotted lines).
Due to the boost the endpoint of this null-polygon will be shifted against the
start, and this
deviation may serve as a measure for the boost-parameter. Of course the
same construction can also be applied to the other cases
(e.g.\ {\bf G4}, Fig.\ 11).}
\end{quote}
\end{figure}

We have so far neglected the boost-component of the generator. After all
the fully-fledged generator of the subgroup will be ${s_1}^nb_\o$!
Does this have any consequences on the factor space?
According to the theorem at the beginning of this section we must check
whether the corresponding subgroups are conjugate.
This is not the case here: 
The generator ${s_1}^n b_\o$ of the subgroup commutes with everything
except flips, and even a flip only transforms the group elements to their
inverse, $f{s_1}^nb_\o f={s_1}^{-n}b_{-\o}=({s_1}^nb_\o)^{-1}$.
Thus the group remains the same, and
one cannot change the boost-parameter $\o$ by conjugation.
Let us again look for some geometrical meaning of this parameter.
In the case of pure boosts discussed above (which also lead to cylinders)
we have found the metric-induced
circumference as such an observable. This cannot be transferred to the
present case, however: there is no closed $X^3=const$-line along which a
circumference could be measured, only the number of blocks $n$ `survives'.
So we have to be  more inventive: one could, e.g.,
take a series of null extremals, zigzagging
around the cylinder between two fixed
values of $X^3$ (see dotted lines in Fig.\ 12)
and interpret the deviation from being closed, i.e.\ the distance between
starting- and endpoint on this $X^3=const$-line, as measure for the boost.
This distance is of course independent of the choice of the starting point on
the $X^3=const$-line, but it depends on the two $X^3$-values; in particular,
these $X^3$-values can always be chosen such that the deviation is zero
(closed polygon). This interpretation of the boost-parameter may be  
somewhat technical (we will find a much nicer one for e.g.\ {\bf G8,9} below), 
but there is certainly no doubt that the parameter is geometrically
significant.

\medskip

\noindent{\bf Two non-degenerate horizons (e.g.\ G8,9, J1):}

\noindent%
Here $\smovs=\langle s_1,s_2 \,|\, {s_1}^2=1, (s_1{s_2}^{-1})^2=1 \rangle$,
where the second relation may be replaced by $s_1s_2s_1={s_2}^{-1}$.
With the help of these two relations any element can be expressed in
the form ${s_2}^n$ or ${s_2}^ns_1$, and the group
can thus (in coincidence with \re{smovsplit})) be written as a semi-direct
product, $\ZPT\ltimes{\Z}$,
where $\ZPT:=\{1,s_1\}$ and ${\Z}:=\{{s_2}^n,n\in{\Z}\}=\osmovs$.
However, not all those elements can be permitted: $s_1$ is a reflexion at one
bifurcation point, $s_2s_1=\scoat$ is a reflexion at the other bifurcation
point, and the general element ${s_2}^ns_1$ is
conjugate to one of them, and thus is also a bifurcation point reflexion.
The only freely acting group elements are
thus the ${s_2}^n$ and they shift the manifold horizontally (in Figs.\
13,14) a number of copies sidewards. The factor space is then a
cylinder. Each cylinder carries again a further real boost-parameter, but
let us postpone this discussion and first admit flips
(i.e.\ non-orientable solutions). 
The most general transformations involving flips (still omitting boosts) are
${s_2}^nf$ and ${s_2}^ns_1f$, but ${s_2}^nf$
and $s_1f$ are pure flips and cannot be used. A small calculation shows
further that the only remaining admissible groups are those generated by {\em
one\/} element ${s_2}^ns_1f$, $n\ne 0$. Since $s_1f$ is a flip at the middle
sector (a reflexion at the horizontal axis in Fig.\ 13) and
${s_2}^n$ is a shift along that axis, the whole transformation is a
glide-reflexion and the factor space a M\"obius-strip of $n$ copies
circumference.

\begin{figure}[htb]
\begin{center}
\leavevmode
\epsfxsize \textwidth \epsfbox{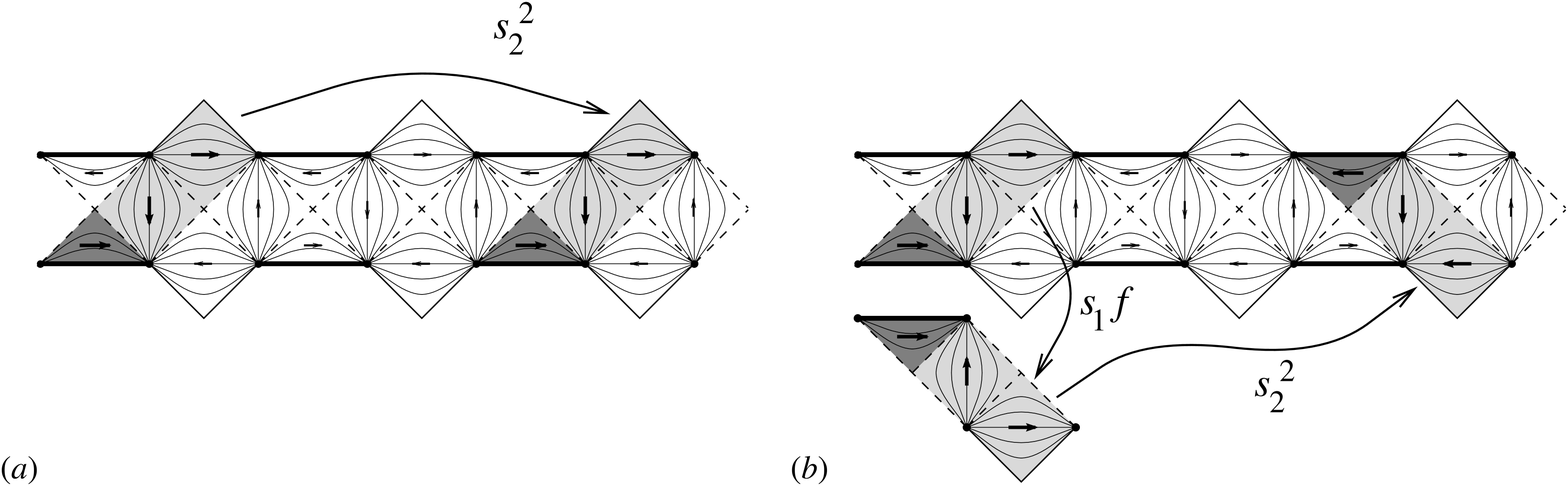}
\end{center}
\renewcommand{\baselinestretch}{.9}
\small \normalsize
\begin{quote}
{\bf Figure 13:} {\small Sector-moves for {\bf G8,9}; ${s_2}^2$ gives
rise to a cylinder ({\it a\/}), while ${s_2}^2s_1f$ yields a M\"obius-strip
({\it b\/}). Note that in ({\it b\/}) the (Killing-)arrows in the blocks
identified by ${s_2}^2$ do not match.}
\end{quote}
\end{figure}

Now concerning the boost-parameter:
The complete cylinder-generator is ${s_2}^nb_\o$. So we must check
whether subgroups with different parameters $\o$ are conjugate.
Of course conjugation with $s_2$, $f$, and boosts does not change the
parameter. However, conjugation with $s_1$ or $s_1f$ changes $\o$ to its
negative, $(s_1f)({s_2}^n b_\o)(s_1f)^{-1}={s_2}^n b_{-\o}$.
Hence the factor spaces
corresponding to $\o$ and $-\o$ are isometric and one could consider to
restrict the boost-parameter to non-negative values, $\o\in{{\R}_0}^+$.
However, the transformation $s_1f$ is a flip at the central sector, i.e., a
reflexion at the longitudinal axis (horizontal in Figs.\ 13, 14)
and thus inverts the time.
Hence, if the spacetime is supposed to have a time-orientation,
then the generators with boost-parameter $\o$ and $-\o$ are no
longer conjugate in the restricted symmetry group and the parameter has to
range over all of $\R$.

In order to give this parameter a geometrical meaning we could of course
employ the `zigzagging' null-polygon again, but in this case there
is a much better description:
In Fig.\ 14\ part of the infinitely
extended solution {\bf G8,9} is drawn. A generating sector-move
shifting the solution two copies (in this example) to the right dictates
that the rightmost large patch (four sectors) is pasted
onto the corresponding left patch. The boost-parameter then describes that
before the pasting a boost has to be applied to the patch. As shown
in Fig.\ 14\ its effect is, e.g., that the thin timelike line crossing
the right patch vertically has to be glued to the curved line of the left
patch. The (shaded) region between these two lines is a possible fundamental
region for this factor space! Also, due to the boost the spacelike tangent
vector (arrow) to the dotted curve is tilted.

\begin{figure}[htb]
\begin{center}
\leavevmode
\epsfxsize 10cm \epsfbox{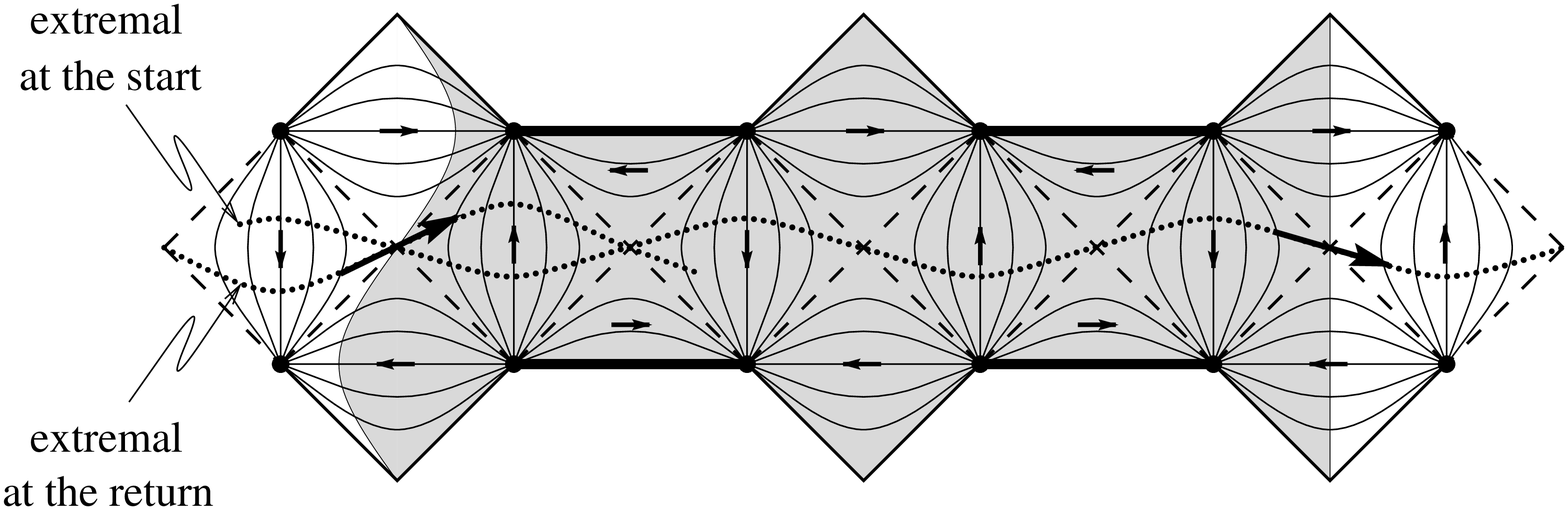}
\end{center}
\renewcommand{\baselinestretch}{.9}
\small \normalsize
\begin{quote}
{\bf Figure 14:} {\small Fundamental region (shaded) and interpretation
  of the boost-parameter for {\bf G8,9}. The right straight boundary has to be
  glued to the curved left one, and also the extremal \re{special})
  (dotted line) returns tilted (i.e.\ boosted).}
\end{quote}
\end{figure}

In {\bf II} it was shown that the bifurcation points are conjugate
points and the extremals running between them are those of \re{special}) (in
Fig.\ 14, they have been drawn as dotted curves). They run through the
bifurcation points into all directions between the two null-directions.
A boost bends them sidewards, altering the angle of their tangent.
One can now start from a bifurcation point in a certain direction
along a spacelike extremal. This extremal will eventually return to the
original point, but due to a boost its tangent (cf.\ arrow in Fig.\ 14)
at the return  may be tilted (boosted) against that at the start. 
This boost is of course independent of the chosen
extremal and is thus a true `observable'; in particular, there is one solution
without a boost. Thus,
the cylindrical solutions are
parametrized by a positive integer (number of patches) and a real constant
parametrizing the boost.% FOOTNOTE
\footnote{If we had chosen $\a <0$ in (\KVref), then the whole Penrose
  diagram would have to be rotated by $90^\circ$.  The above extremals
  would then be timelike and the boost at the return could be
  interpreted nicely as acceleration during one journey around the
  cylinder.}
As discussed before, if there is no time-orientation, then the
boost-parameter has to be restricted to ${{\R}_0}^+$.
In the above interpretation of the boost-parameter this restriction
arises if one cannot distinguish
between boosts to the past or future. Of course, if a time-orientation is
given, then such a distinction is possible.
Note that in this case the
(time-)\- direction of the boost is independent of the sense
in which the extremal runs through the diagram.

It may seem that for the M\"obius-strip one would also have such a continuous
parameter due to different flips. However, this is {\em not\/} the case:
Any two flips are conjugate
(via a boost, $f'\equiv fb_\o=b_{-\o/2}fb_{\o/2}\sim f$), hence the
corresponding subgroups are conjugate and all 
M\"obius-strips are equivalent. 
(There is always one extremal \re{special}) which returns unboosted).
Also, in contrast with the former examples,
a boost cannot be defined consistently on the M\"obius-strip;
the boost transformation does not `factor through' the canonical projection
onto the factor space. This is also seen immediately from Fig.\ 13
({\it b\/}), where in the M\"obius-case the sectors
occasionally have to be identified with their mirror images and thus the
arrows indicating the boost-direction do not match.
However, locally this Killing symmetry is still
present.
Hence, the M\"obius-strip solution is only parametrized by a
positive integer (number of copies).

\section{More than one generator}
\plabel{Two}

So far we have treated the cases where $\osmovs\cong{\Z}$ (one
generator) or trivial. 
In those cases also all
subgroups have been one-generator groups $\Z$, and the possible
topologies have thus been restricted to cylinders and M\"obius strips
(remember that the subgroup $\cal H$ factored out equals the fundamental group
of the factor space, $\p_1({\cal M}/{\cal H})$).
This situation changes drastically when there is more than one generator,
as there are then subgroups of arbitrarily high rank (even infinite).

Ultimately we want to know the conjugacy classes of subgroups of \osmovs\
(in this section we restrict ourselves to space- and time-orientable
solutions; then all those subgroups are properly
acting). Subgroups of free groups are again free, so in principle any
solution can be obtained by the choice of a free set of
generators. But this is only the easier part of the job:
\begin{itemize}
 \item Given a subgroup (say, in terms of generators) it may be hard to find a
  {\em free\/} set generating this group.
 \item Also the free generators are by no means unique (one can, e.g., replace
  $g_1$, $g_2$, \ldots by $g_1$, $g_2g_1$, \ldots). Only the number of free
  generators (the {\em rank\/} of the subgroup) is fixed. So there is the
  problem to decide whether two sets of generators describe the same group
  or not.
 \item We have to combine the subgroups into conjugacy classes.
\end{itemize}
Since the group is free, these three issues can be solved explicitly (at
least for finitely generated subgroups);% FOOTNOTE
\footnote{Surprisingly, for non-free groups this is in general impossible.
  For instance, there is no (general) way to tell whether two given words
  represent the same group-element (or conjugate elements); and it may also
  be undecidable whether two presentations describe isomorphic groups
  (word-, conjugation-, and isomorphism-problem for
  combinatorial groups, cf.\ \cite{CombGr}).}
however, the algorithms are rather cumbersome and thus we will not extend
this idea here (details in \cite{CombGr}).

\medskip

Due to the more complicated fundamental groups it is to be expected
that one gets interesting topologies.
As already mentioned in Sec.\ \ref{Intro}, all the solutions will be
non-compact (this is also clear, since there is no compact manifold without
boundary with a free fundamental group!).
Thus it would be nice to have a classification of non-compact
surfaces at hand. Unfortunately, however, there is no really satisfactory
classification which could be used here (cf.\ \cite{Massey}).
Let us shortly point out the wealth of different possibilities:
A lot of non-compact surfaces can be obtained by cutting holes into compact
ones. Of course the number of holes may be infinite, even uncountable
(e.g.\ a Cantor set). A more involved example is that of surfaces of
countably infinite genus (number of handles). Finally, there need not even be
a countable basis of the topology (this does not happen here, though, since by
construction there are only countably many building blocks involved).
For an important subcase (finite index), however, the resulting topologies
are always of the simple form `surface of finite genus with finitely many
holes'.

Abstractly the index of a subgroup $\cal H$ is the number of cosets of
$\cal H$. But it has also a nice geometrical meaning:
Since \smovs\ acts freely and transitively on the sectors of the same type
(and thus on the building blocks), the index of $\cal H$ in \smovs\
counts the number of building blocks in the fundamental region. Actually, 
since we started from \osmovs, it would be more convenient to use
the index of $\cal H$ in that group. If all horizons are of even degree,
then $\smovs=\osmovs$ and there is thus no difference.
On the other hand, if there are horizons of odd degree, then $\osmovs$ is a
subgroup of index 2 of $\smovs$ and consequently the number of building
blocks is {\em twice\/} the index of $\cal H$ in \osmovs. This is also obvious
geometrically, since the fundamental regions in this case are built from
patches consisting of two building blocks (e.g.\ those situated
around a saddle-point in the examples of  Figs.\ 15\
({\it b\/}), 16\ below).
The index counts the number of these basic patches in the 
fundamental region then. For finite index, furthermore, it is correlated
directly to the rank of $\cal H$ via formula \re{index}) (with
$\cal G$ replaced by \osmovs), thus
\begin{equation}
   \mbox{index}\, {\cal H}= 
   \mbox{number of basic patches} = \frac{\mbox{rank}\,{\cal H}-1}{n-1}\,
  \plabel{indexH}
\end{equation}
($n$ being the number of generators of \osmovs).

We will now provide the announced examples, starting with a
discussion of the respective combinatorial part of the
(orientation and time-orientation-preserving) symmetry group and
followed by a discussion of possible factor spaces. Figures
15\ and 16\ contain basic patches as well
as the generators of \osmovs.  Although this group is the same
in all these cases (rank 2), its action for {\bf R5} is
different from the others, and correspondingly will be found to
give rise to different factor solutions.

\begin{figure}[htb]
\begin{center}
\leavevmode
\epsfxsize 10cm \epsfbox{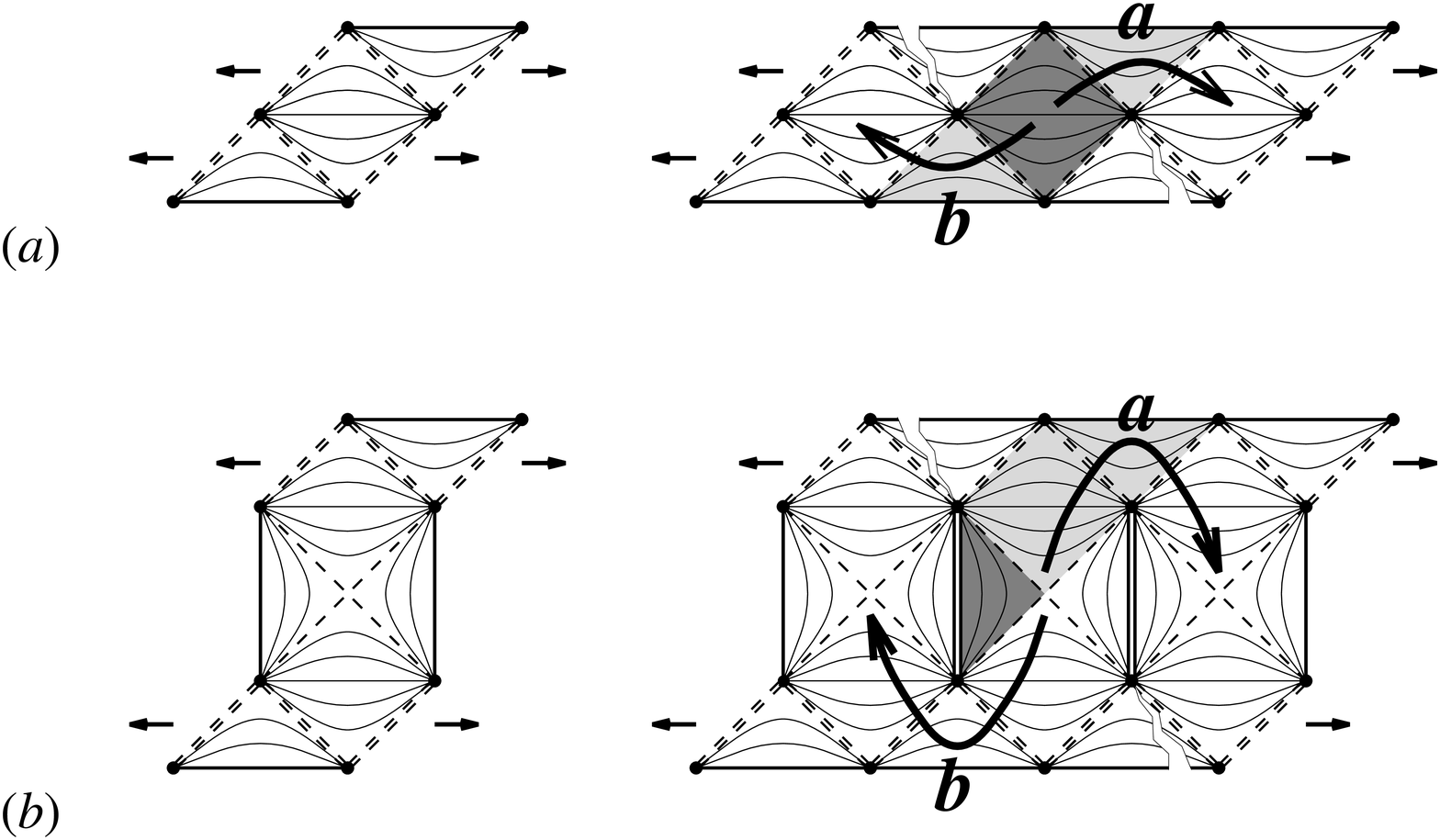}
\end{center}
\renewcommand{\baselinestretch}{.9}
\small \normalsize
\begin{quote}
{\bf Figure 15:} {\small Basic patches and
generators of \osmovs\ for a fictitious
example with two doubly degenerate horizons ({\it a\/}) and for {\bf R3}
({\it b\/}). Note that in the upper example the points in half height are at
infinite distance, and
that in the lower example the vertical singularities meet, yielding a
slit (double lines). Going once
around this point/slit leads into a different layer of the
universal covering, as indicated by the jagged lines (multilayered Penrose 
diagrams). Consequently $ab \neq 1$.}
\end{quote}
\end{figure}

In the example Fig.\ 15\ ({\it a\/}) (two doubly degenerate horizons)
the group \smovs\ ($=\osmovs$) is free already, with the
two generators $s_1$ and $s_2$. Geometrically,
however, the moves with basis-sector 1 have a better representation.
Let $a:=\scoat \equiv s_2{s_1}^{-1}$ and $b:=\scoaz \equiv {s_1}^{-1}$.
Clearly $a$ is a move one block to the right {\em above\/} the singularity
and $b$ a move to the left {\em below\/} the singularity. Note that
since we are in the universal covering their composition, $ab$, is {\em
not\/} the identity but leads into another layer of the covering;
if an identification is to be enforced,
then the element $ab$ must occur in the factored out subgroup.

In Fig.\ 15\ ({\it b\/}) only the second horizon is degenerate
{\bf (G7,10, R3,4)}. Thus the basic patch consists of two building blocks.
Here \osmovs\ is a proper subgroup of \smovs\ and has the two free
generators $s_1s_2$ and $s_2s_1$ (since the
second sector is spatially homogeneous but the basis-sector stationary). 
Again, $a:=s_2s_1$ is a move one patch to the right {\em above\/} the
singularity and $b:=s_1s_2$ a move to the left {\em below\/} the singularity. 
Thus the action is similar to that in Fig.\ 15\ ({\it a\/}).
However, if (time-)\-orientation-preservation is not
required, then here one has the additional symmetries $f$ and $s_1$, i.e.\
reflexion at the horizontal axis or at the saddle-point respectively.

Our last example is {\bf R5, G11} (three non-degenerate horizons,
cf.\ Figs.\ 16, 2). Also in
this case there are two free generators: $a:=s_2$, which is a move one
patch upwards, and $b:=s_3s_1$, a move one patch to the right. (The other
two potential generators can be expressed in terms of $a$ and $b$ by means
of the saddle-point relations: $s_1s_2s_1=a^{-1}$ and
$s_1s_3=a^{-1}b^{-1}a$). Again, going once around the
singularities leads into a new layer of the universal covering
($\LRA [a,b]:=aba^{-1}b^{-1}\ne 1$).

\begin{figure}[htb]
\begin{center}
\leavevmode
\epsfxsize 10cm \epsfbox{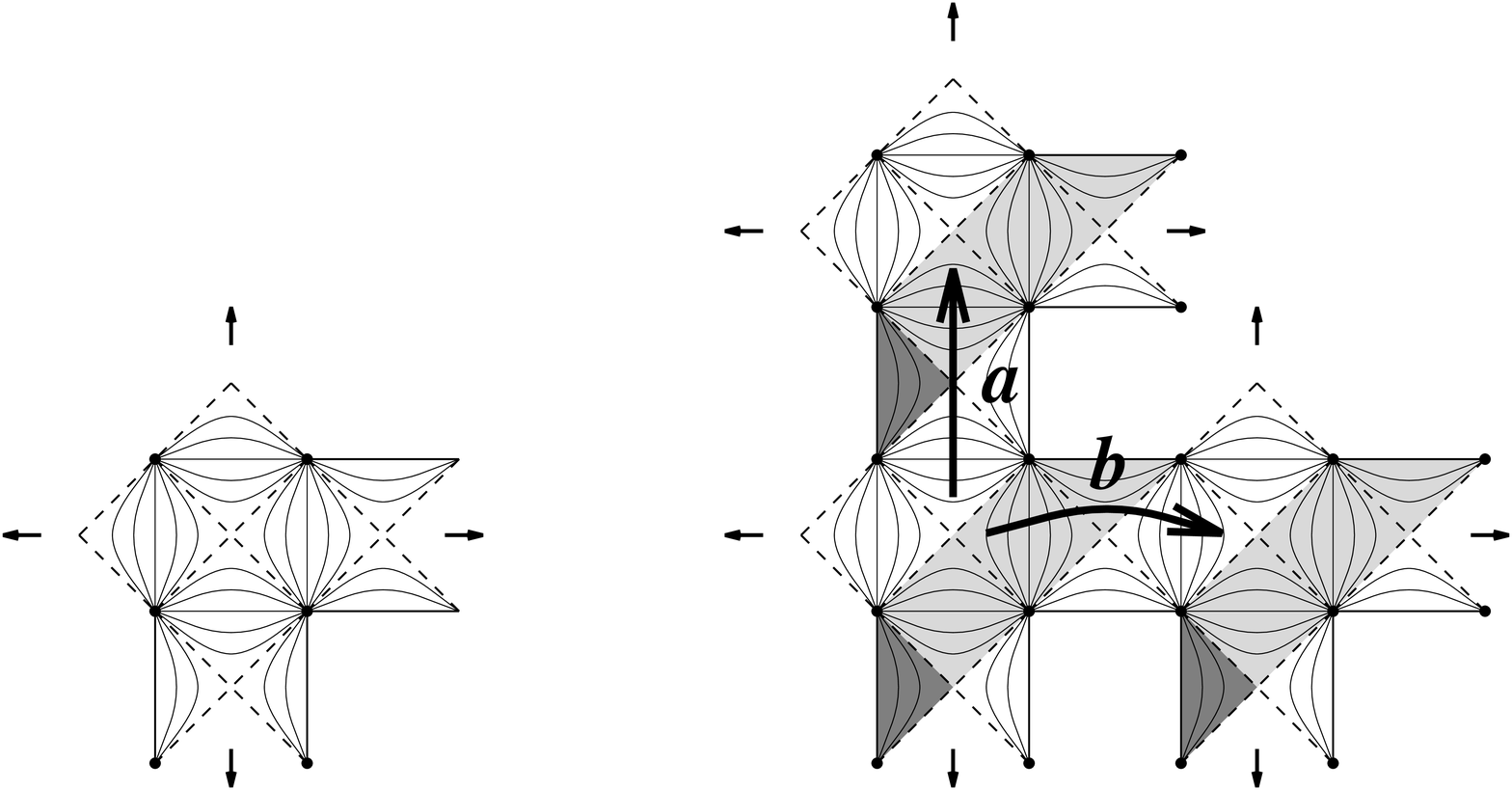}
\end{center}
\renewcommand{\baselinestretch}{.9}
\small \normalsize
\begin{quote}
{\bf Figure 16:} {\small Possible basic patch and generators of 
\osmovs\ for {\bf R5}. As is seen here, the basic patch need not
consist of two {\em entire\/} blocks, but the involved sectors may be
rearranged somewhat.}
\end{quote}
\end{figure}

Let us now determine the topology for the solutions with finite index.
It is clear that then there are also only finitely many boundary segments.
[This is a slightly informal terminology, since these `boundary
segments' (singularities, null infinities, points at an infinite distance)
do {\em not\/} belong to the manifold. Still, this can be made precise and
such boundaries are called `ideal boundaries' or `ends'; we will
thus simply use the notion boundary.] 
The generators of the subgroup $\cal H$ determine
how the faces of the fundamental region have to be glued and thus also how
the boundary segments are put together to form boundary components.
This is shown as two examples in Fig.\ 17. There opposite faces
should be glued together, which can be achieved by using the following
generators: $b$, $a^{-1}ba$, $a^{-2}ba^2$, $a^3$ for the left case,
and  $b$, $a^{-1}b^2a$, $a^{-1}bab^{-1}a$, $a^2$ in the right case, provided
one starts from the lowest basic patch. When starting from another
patch, the subgroups and their generators will be conjugates of the above
ones, but clearly this does not change the factor solution.

Now, topologically to each boundary component (which is clearly an $S^1$) 
a disk can be glued. This yields a compact orientable surface,
which is completely determined by its genus. The original manifold is then
simply this surface with as many holes as disks had been inserted (each
boundary component represents a hole).
The genus in turn depends on
the rank of the fundamental group $\p_1({\cal M}/{\cal H})\cong{\cal H}$
and on the number of holes:
\begin{equation}
   \mbox{rank}\, {\cal H}= 
   \mbox{rank}\,\,\p_1({\cal M}/{\cal H}) = 2\,\, \mbox{genus} +
                                             \mbox{(number of holes)}-1\,.
  \plabel{rankH}
\end{equation}
Expressing the rank by means of \re{indexH}), this yields
\begin{equation}
 \mbox{genus}\,=\,\frac{(\mbox{number of basic patches})\cdot(n-1)-
                     (\mbox{number of holes})}2+1 \,,
  \plabel{genus}
\end{equation}
where $n$ is the number of generators of \osmovs.

The general procedure to determine the topologies of the factor spaces can
thus be summarized as follows:
Draw a fundamental region for your chosen subgroup and
determine which faces have to be glued together. Then count the connected
boundary components (= number of holes) and calculate the genus from
\re{genus}). We illustrate this procedure at two examples in Fig.\ 
17. 
\begin{figure}[htb]
\begin{center}
\leavevmode
\epsfxsize 10 cm \epsfbox{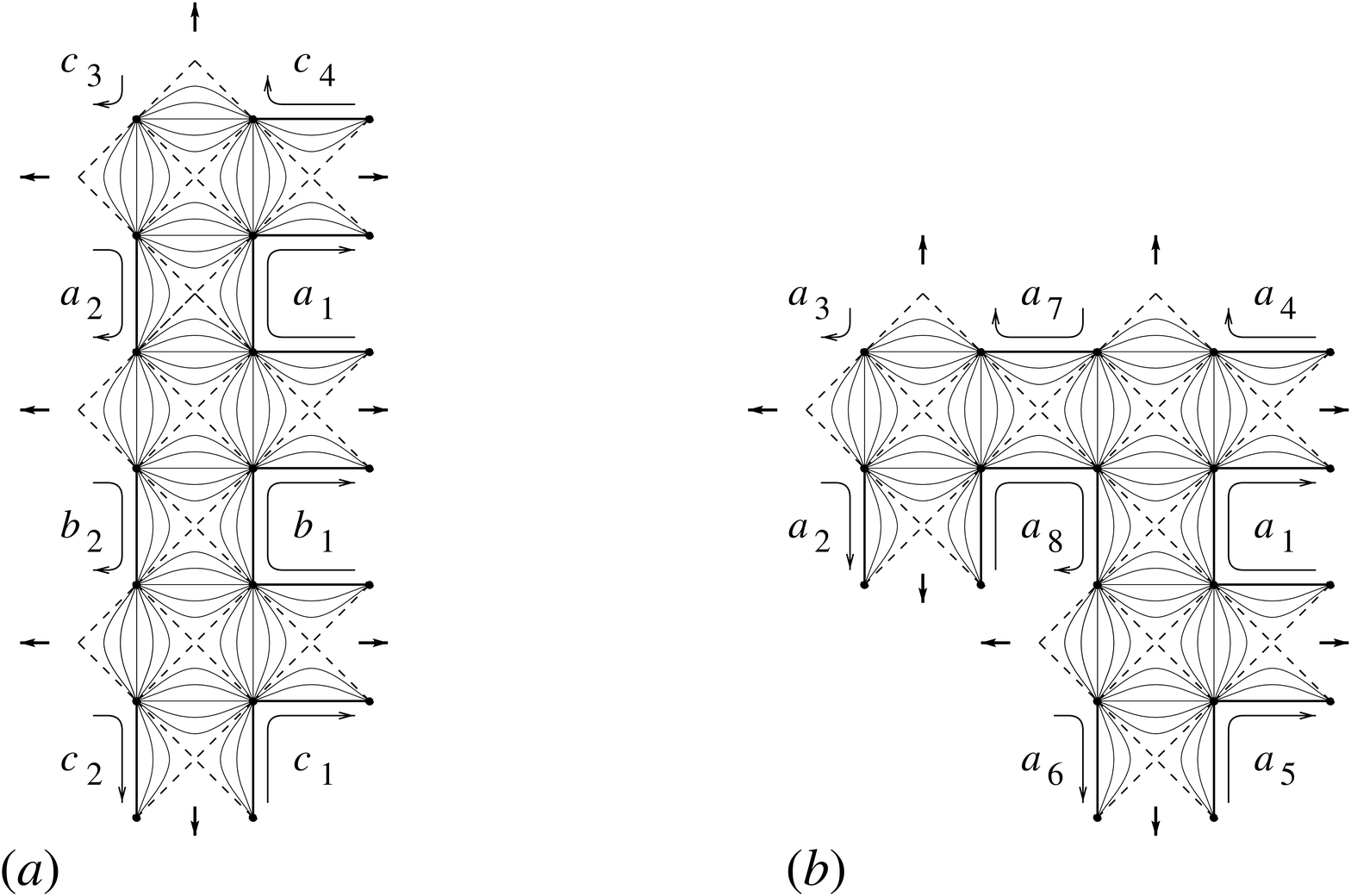}
\end{center}
\begin{quote}
{\bf Figure 17:} {\small Counting the boundary components
(opposite faces have to be glued together). In both cases there are three
basic patches (cf.\ left part of Fig.\ 16) 
and thus (use  \re{index}) and $\mbox{rank}\,\osmovs = 2$ !) 
four generators for $\cal H$, given in the text. 
However, due to the different number of boundary components (holes) the
topologies differ, cf.\ Eq.\ \re{genus}): In the
left example there are three components ($a_{1,2}$, $b_{1,2}$,
$c_{1\!-4}$), thus the topology
is that of a torus with three holes. In the right example there is only
one component ($a_{1\!-8}$); therefore this solution is a
genus-2-surface with one hole. The resulting manifolds are shown in
Fig.\ 5.} 
\end{quote}
\end{figure}
Some further examples for {\bf R5} are given in Fig.\ 18. 
Actually, they show that surfaces of any genus ($\ge1$) and with any number
($\ge1$) of holes can be
obtained: continuing the series Fig.\ 18\ ({\it b\/}),
({\it c\/}), ({\it d\/}) one can increase the number of handles arbitrarily,
while attaching single patches from `below' like in ({\it e\/})
allows one to add arbitrarily many holes.

\begin{figure}[htb]
\begin{center}
\leavevmode
\epsfxsize \textwidth \epsfbox{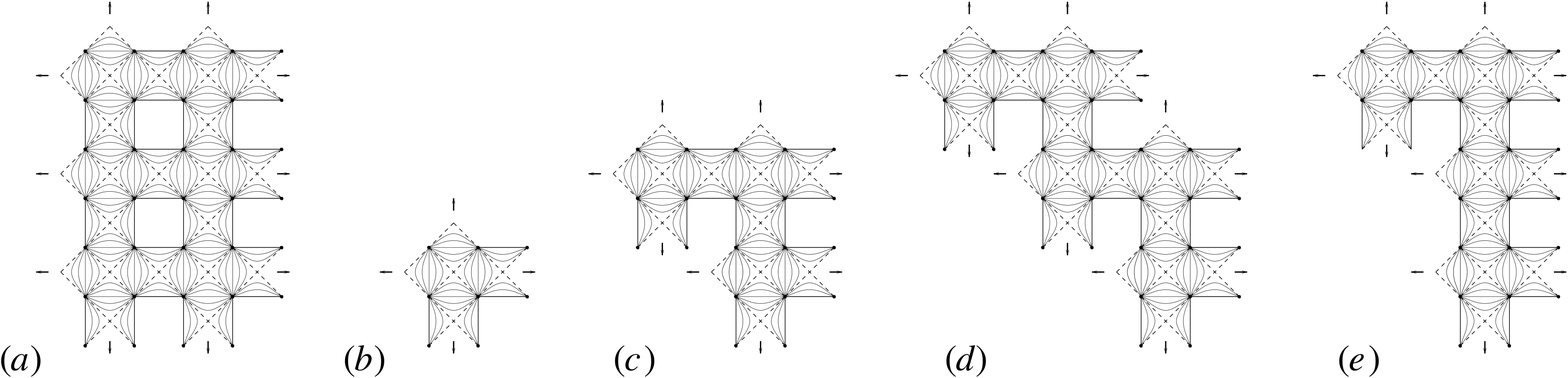}
\end{center}
\renewcommand{\baselinestretch}{.9}
\small \normalsize
\begin{quote}
{\bf Figure 18:} {\small Fundamental regions of factor spaces for
{\bf R5}. Opposite faces have to be glued together.
({\it a\/}) torus with six holes,
({\it b\/}) torus with one hole,
({\it c\/}) genus-2-surface with hole,
({\it d\/}) genus-3-surface with hole,
({\it e\/}) genus-2-surface with two holes.}
\end{quote}
\end{figure}

For the cases of Fig.\ 15\ (e.g.\ {\bf R3}) the same analysis can
be applied. For instance, it is obvious that cylinders with an arbitrary
number of holes ($\ge 1$) can be obtained. Also surfaces of higher genus are
possible; however, now the number of holes is always $\ge3$ (past and future
singularity, and at least one hole in `middle height'). [Note that this is
no contradiction to the cylinder-with-hole case, since a cylinder with one
hole is a sphere (genus-0-surface) with three holes].
In contrast to the {\bf R5}-examples these factor solutions
do not have closed timelike curves.

\medskip

We turn to the cases with infinite index and thus infinite 
fundamental regions. All subgroups of infinite rank belong to this 
category, but also many subgroups of finite rank (see below). One topological
reason for an infinite rank of the subgroup (and thus also of the fundamental
group $\p_1$) is the occurrence of infinitely many holes. For instance, it was
pointed out that in the solution {\bf R3} (Fig.\ 15) the move $ab$ is
not the identity but leads into a new layer of the universal covering. One
can of course enforce the identification of overlapping layers by imposing
the relation $ab\stackrel!=1$ and its consequences. This is tantamount to
factoring out the group generated by $ab$ and all its conjugates
(the elements $a^kaba^{-k}$, $k\in{\Z}$,  form already a free set of
generators). The result is a ribbon with infinitely many holes (slits). 
Clearly the parameter space of such solutions is
infinite dimensional now (cf.\ also remarks at the end of this section).  
[If  in addition one imposed the relation
$a^n\stackrel!=1$, then the previously infinite set of
generators would boil down to $n+1$ generators ($a^n$, and
$a^kaba^{-k}$ for $0\le k<n$), and the resulting factor space
(finite index again) would be a cylinder with $n$ holes.]

Likewise, in the example {\bf R5} (Fig.\ 16) the identification of
overlapping layers is obtained by factoring out the infinitely generated
commutator subgroup (generated freely e.g.\ by $a^mb^nab^{-n}a^{-m-1}$,
$(m,n)\ne(0,0)$; cf.\ \cite{Massey}); the factor space is a planar,
double-periodic `carpet' then.
Adding, furthermore, the generator $a^n$ (or $b^n$) yields a cylinder with
infinitely many holes (e.g.\ Fig.\ 18\ ({\it a\/}) extended infinitely 
in vertical direction); and adding both $a^n$ and $b^k$ yields a torus with
$nk$ holes, which is again of finite index.

Another possible reason for an infinite rank is an infinite genus (number of
handles); such a solution is obtained for instance by  continuing
the series Fig.\ 18\ ({\it b\/}), ({\it c\/}), ({\it d\/})
infinitely. Of course, both cases can occur
simultaneously (infinite number of holes {\em and\/} infinite genus).

Let us now discuss the groups of finite rank and infinite index.
Already the universal covering itself, being  topologically an open
disk (or $\R^2$),% FOOTNOTE
\footnote{The reader who has difficulties in imagining that such an 
  infinitely branching patch is really homeomorphic to a disk may
  recall the famous Riemann mapping theorem, which states that
  any simply connected (proper) open subset of $\R^2$, however
  fractal its frontier might be, is not only homeomorphic but even
  biholomorphically equivalent to the open unit disk
  (e.g.\ \cite{Burckel}). [Clearly, the universal covering is a priori
  not a subset of the plane (due to the overlapping layers), but
  by a simple homeomorphism it may be brought into this form.]}
provides such an example
(with the trivial subgroup factored out). 
But also one-generator subgroups can by \re{index}) never be of finite
index (if $\mbox{rank}\osmovs\ge2$).
These subgroups yield proper (yet slightly pathological) cylinders without
holes: Let ${\cal H}_g=\langle g \rangle \equiv\{g^n,n\in{\Z}\}$ for arbitrary
non-trivial $g \in \osmovs$.
In a graphical form this may be thought of as that
the generator $g$ of this infinitely cyclic
subgroup% FOOTNOTE
\footnote{or rather the corresponding cyclically reduced element
  (remember that conjugate groups yield isomorphic factor spaces).}
defines a path in the universal covering. Now the end-sectors of the path
(i.e.\ of the corresponding ribbon) are identified and at all other junctions
the solution is extended infinitely without further identifications (cf.\
Fig.\ 19). Thus a topological cylinder, although with a
terribly frazzled boundary, is obtained (as before it is possible to
smooth out the boundary by a homeomorphism). [The cylinders obtained in
this way may have kinks of the lightcone or not. For instance, in the example
Fig.\ 19\ the lightcone tilts by (about) $90^\circ$ and then tilts
{\em back\/} again. So, this is not a kink in the usual sense of the word;
still, there is not one purely
spacelike or purely timelike loop on such a cylinder.]

\begin{figure}[htb]
\begin{center}
\leavevmode
\epsfxsize 6cm \epsfbox{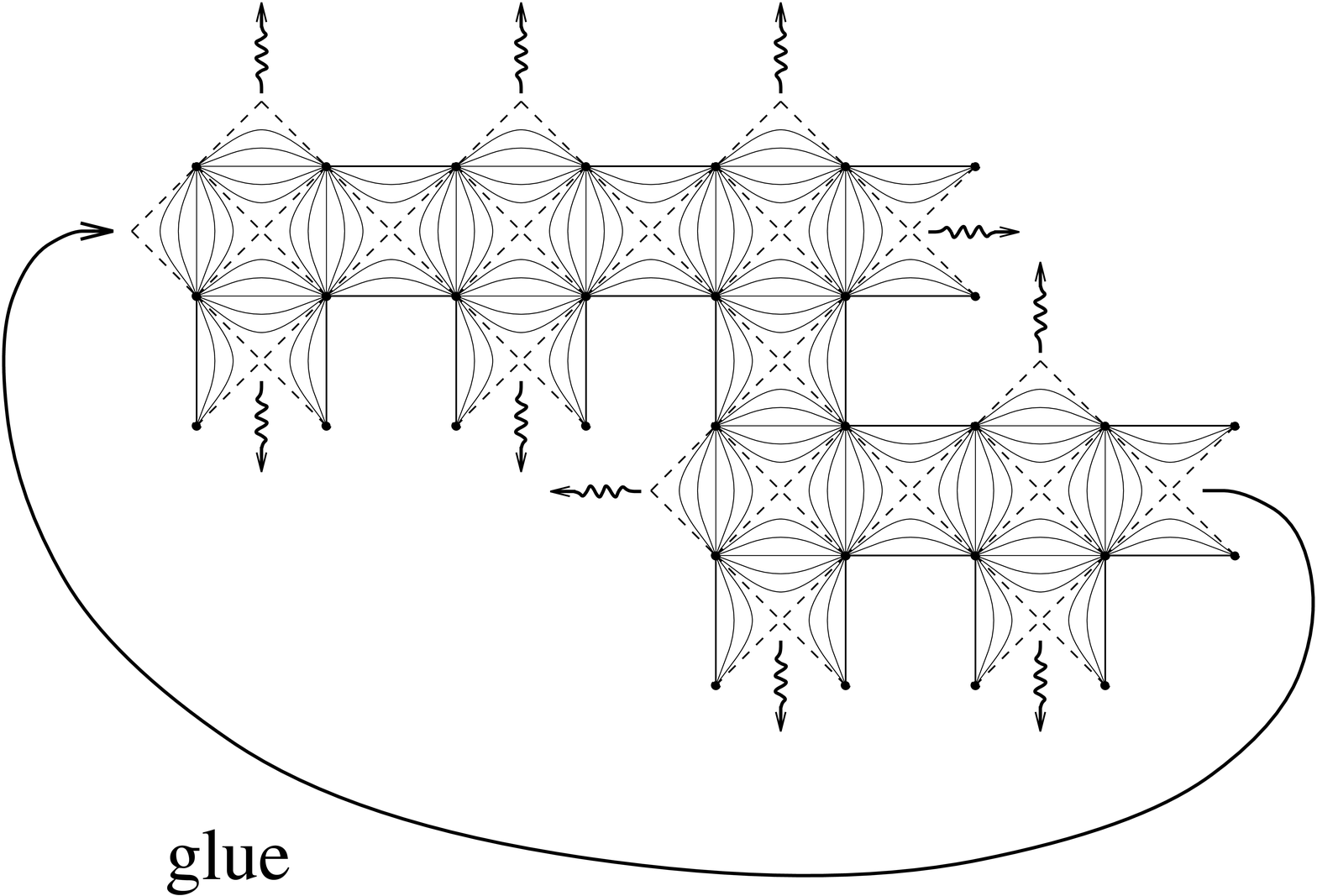}
\end{center}
\renewcommand{\baselinestretch}{.9}
\small \normalsize
\begin{quote}
{\bf Figure 19:} {\small Frazzled cylinder from {\bf R5}. Only the
utmost left and right faces have to be glued together, such as to make a
closed `ribbon' of five patches (a possible generator for this gluing is
e.g.\ $b^2a^{-1}b^2$). At all other faces (indicated 
$\rightsquigarrow$) the solution has to be extended without further
identifications, similarly to the universal covering.}
\end{quote}
\end{figure}

And even for higher ranks of $\cal H$ there are solutions of infinite index.
The topologies obtained in this way are again of the simple form compact
surface with hole(s). There is, however, a much greater flexibility
in the rank (which is no longer restricted by formula \re{indexH}))
as well as in the number of boundary-components (remember that in the examples
Fig.\ 15 all solutions of finite index had at least
three boundary-components). Indeed, one can obtain any genus and any
nonzero number of holes in this way, as described briefly in the following
paragraph. Eq.\ \re{rankH}) is still valid, however, $\mbox{rank}\, {\cal H}$
can no longer be expressed by \re{indexH}) but has to be determined directly
from the gluings.

As already mentioned we may abuse the classification of compact 2-manifolds
with boundary for our purpose; one just has to replace the true boundaries
by `ideal boundaries' which do not belong to the manifold. 
Note that the periphery of the basic patch consists of a couple of faces
which have to be glued together pairwise, separated by (`ideal') boundary
components. [It is topologically immaterial whether these boundary components
are pointlike or extended singularities, since they do {\em not\/} belong to
the manifold; each boundary point can be stretched to an extended segment by a
homeomorphism and vice-versa.]
The faces come in pairs and the number of pairs equals the rank of the group
\osmovs. Thus in the present case there are at least two such pairs, which is
sufficient to produce fundamental regions with arbitrarily many faces.
Furthermore, by virtue of the infinitely branching extensions one can get
rid of redundant faces: just extend the solution infinitely at this face
so as to obtain a new (`frazzled') boundary segment which connects the two
adjacent ones, yielding one larger boundary segment. Thus it is possible to
produce polygons, the faces of which are to be glued in an arbitrary order.
According to \cite{Massey} this already suffices to produce all topologies
announced above.

\medskip

Finally, we have to discuss the boost-parameters.
Since each free generator of $\cal H$ carries a boost-parameter, their total
number equals the rank $r$ of this subgroup. 
Also an interpretation can be given in analogy to the cases dealt with
before (zigzagging null-polygon and/or boosted saddle-point extremals).
However, not all such choices of 
an $r$-tuple of real numbers are inequivalent.
We show this with the example of the torus with three holes (Fig.\ 17, 
left part): There are four generators and thus also four
boost-parameters, three of them describing the freedom in the
horizontal gluing ($b$, $a^{-1}ba$, and $a^{-2}ba^2$) and one for
the vertical gluing ($a^3$); let us denote them by $(\o_1,\o_2,\o_3;\o_4)$.
Now, since conjugate subgroups lead to equivalent factor spaces, we can
e.g.\ conjugate all generators with $a$. This leads to new generators,
but since $a$ lies in the normalizer% FOOTNOTE
\footnote{The {\em normalizer\/} of a subgroup $\cal H$ contains all
  elements $g$ for which $g^{-1}{\cal H}g = {\cal H}$.}
$\cal NH$ of $\cal H$ in \osmovs\
they still span the same (projected) subgroup  $\cal H$. Thus it is possible
to express the old generators in terms of the new ones.
However, during this procedure the boost-parameters change: For instance,
the (full)% FOOTNOTE
\footnote{Here we denoted the boost %{\em itself\/}
 by $\o_i$ instead of $b_{\o_i}$ in order to avoid confusion with the
 sector-move $b$.}
first generator $\o_1b$ is mapped to $a^{-1}\o_1ba=\o_1(a^{-1}ba)$, i.e.\
the parameter $\o_1$ is shifted from the first to the second
generator. Altogether, the three `horizontal' boost-parameters $\o_{1\!-3}$
are cyclically permuted and thus we get an equivalence relation among the
4-tuples, $(\o_1,\o_2,\o_3;\o_4) \sim (\o_3,\o_1,\o_2;\o_4)$.
This is of course also geometrically evident: while the timelike loop
corresponding to $\o_4$ is uniquely characterized, the three spacelike loops
corresponding to $\o_{1\!-3}$ are indistinguishable (there is no `first
one').

In general, we have a (not necessarily effective) action of
the group ${\cal NH}/{\cal H}$ on the space of boost-parameters
${\R}^r$.  Here $\cal NH$ is the normalizer 
of $\cal H$ in \osmovs\ (or, if no \mbox{(time-)}orientation
is present, also in
\smovs\ or $\Zflip\ltimes\smovs$, respectively).  The true
parameter space is the factor space under this action,
${\R}^{r}\big/({\cal NH}/{\cal H})$. Locally, it is still $r$-dimensional;
however, since the action may have fixed points (e.g. in the above example
the whole plane $(\o,\o,\o;\o_4)$), it is an {\em orbifold\/} only.

\section{Remarks on the constant curvature case}
\plabel{Const}

So far we have only dealt with those solutions where the metric (or
$X^3$-preservation) restricted us to only one Killing field.
For reasons of completeness one should
treat also the (anti-)\-deSitter solutions (\deSitterref) of the
general model, corresponding to the critical values $X^3={X^3}_{crit}$ and
$X^a=0$ (cf.\ {\bf I}), which have constant curvature (and zero torsion).
Constant curvature manifolds have already occurred as
solutions of the Jackiw-Teitelboim (JT) model \cite{JT}, Eq.\ 
(\JTref),  where, however, the symmetry group was still restricted 
to only one Killing field since
$X^3$ had to be preserved. Here, on the other hand, these fields are constant
all over the spacetime
manifold, and thus the solutions have a much higher symmetry. 

Already the flat case offers numerous possibilities: The symmetry group
(1+1 dimensional Poincar\'e group) is generated by translations, boosts, and
if space- and/or time-orientation need not be preserved, also by spatial
and/or time inversion.
As before pure reflexions would yield a boundary line (the reflexion axis)
or a conical singularity (at the reflexion centre) and boosts a
Taub-NUT space. The only fixed-point-free transformations are thus
translations and glide-reflexions% FOOTNOTE
\footnote{A {\em glide-reflexion\/} is a translation followed by a reflexion
  at the (non-null) translation axis. Note that this axis must not be null,
  if the (orthogonal) reflexion is to be well-defined.}
We have thus the following generators and corresponding factor spaces:
\begin{itemize}
\item One translation: Cylinders; parameters = length squared of the generating
translation = circumference (squared) of the resulting cylinder (in
particular, there is only {\em one\/} cylinder with null circumference).
\item One glide-reflexion:  M\"obius-strips, again parametrized by their
circumference.
\item Two translations: Torus, labelled by three parameters: The lengths
(squared) of
the two generators $\vec{a}$ and $\vec{b}$ and their inner product.
Globally, however, this is an overparametrization: Replacing, for instance,
the translation vector $\vec{b}$ by $\vec{b}+n\vec{a}$ changes one length and
the inner product, but still yields the same torus (just the
original longitude is now twisted $n$ times around the torus).
\item One translation and one glide-reflexion: Klein bottle.  Here
  only two parameters survive (the inner product of the two generators
  can be conjugated away always). [This is somewhat similar to the
  situation cylinders versus M\"obius strips in the case of {\bf G8,9} in
  section \ref{Factor}, where a potential continuous parameter
  did not occur due to the non-orientability.]
\end{itemize}
Since two glide-reflexions combine to give a point-reflexion or boost,
and three translations generically yield a non-discrete orbit, this
exhausts all cases.

\begin{figure}[htb]
\begin{center}
\leavevmode
\epsfxsize 9cm \epsfbox{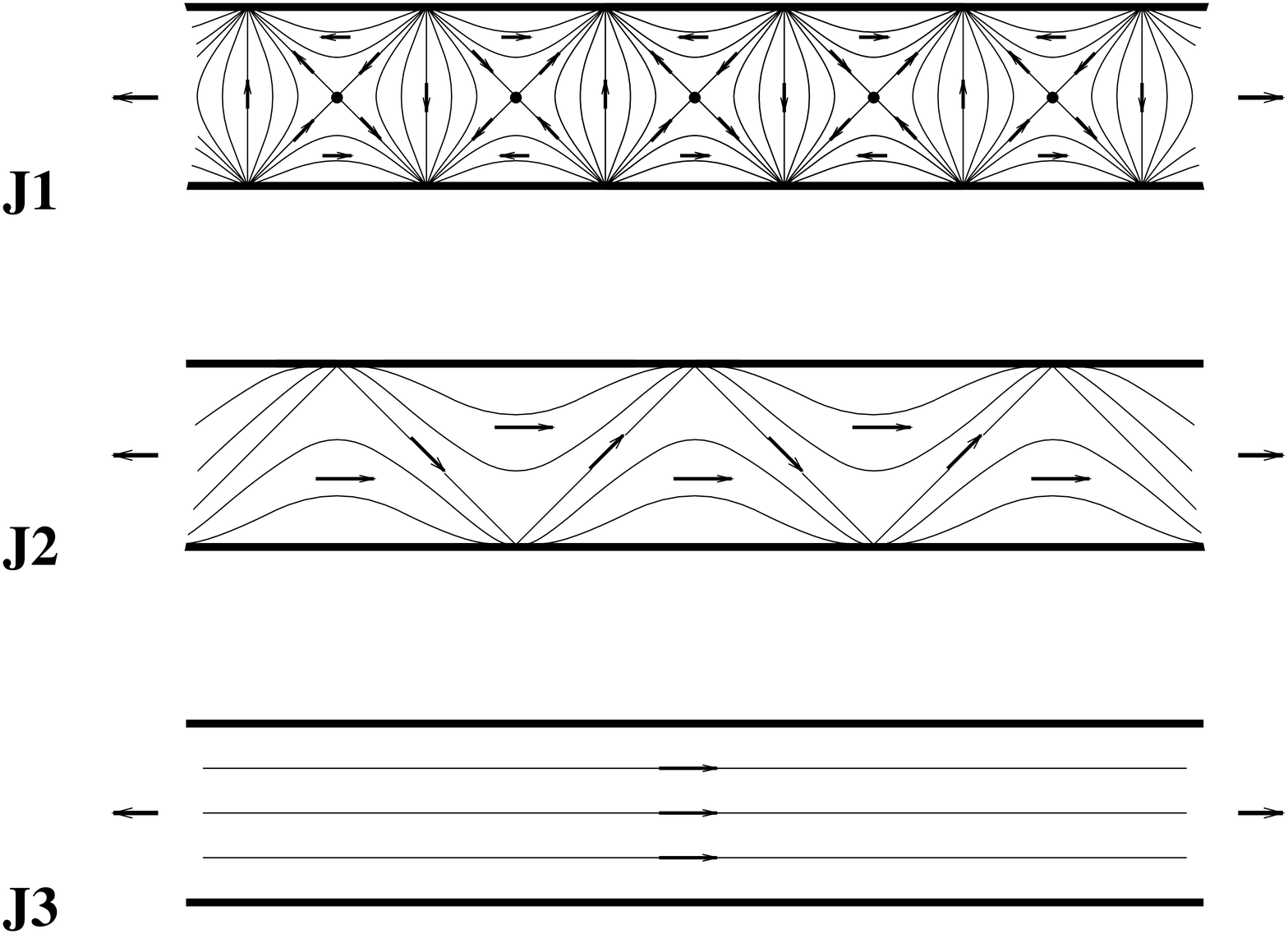}
\end{center}
\renewcommand{\baselinestretch}{.9}
\small \normalsize
\begin{quote}
{\bf Figure 20:} {\small Killing fields for the JT-model
and deSitter.}
\end{quote}
\end{figure}

Now concerning the `proper' ($R\ne0$) (anti-)\-deSitter 
solutions resp.\ their universal coverings:
Here the situation is slightly more involved.
Of course all factor solutions of the JT-model (\JTref)
are also available for the (anti-)\-deSitter case, as both have 
constant curvature and zero torsion. 
For instance, the Killing fields corresponding to the solutions {\bf J1,2}
give rise to cylinders labelled by the block number and a boost-parameter,
while {\bf J3} yields cylinders labelled by their (real number)
circumference. Now, however, there are three independent Killing fields
(e.g.\ {\bf J1,3} and their Lie-bracket)% FOOTNOTE
\footnote{Note that this is the {\em only\/} case with more than one (local)
  Killing field: Whenever curvature is not constant, the Killing trajectories
  are restricted to the lines of constant curvature, which leaves at most one
  independent field. There is thus no 2D-metric with only two local
  Killing fields.}
and thus one would expect further
factor solutions. As an example, one has now in addition to cylinders also
M\"obius-strips of {\em arbitrary\/} circumference and not only of an
integer number of blocks as in the case of {\bf J1 (G8,9)}.
Unfortunately, the full isometry group of the space is $\widetilde O(2,1)$,
whose connected component equals $\widetilde{SL}(2,\R)$, and this group is
famous for having no faithful matrix representation and is thus
rather difficult to handle. A partial classification has been
accomplished by Wolf \cite{Wolf}, who deals with homogeneous spaces% FOOTNOTE
\footnote{A space is called {\em homogeneous\/}, if its group of
  isometries
  acts transitively on it (the space then looks `the same' from every point).}
only and obtains a discrete series of cylinders and M\"obius-strips for them.
There are strong hints that even in the general case these are the
only possible topologies% FOOTNOTE
\footnote{It is clear that no compact topologies can occur: According to
  \cite{Thm} compact Lorentz-manifolds should have an Euler characteristic
  which is zero
  (i.e., torus or Klein bottle), but by the Gauss-Bonnet theorem this is
  impossible for non-vanishing constant curvature.}
(for instance, it would suffice to show that all
properly acting subgroups of $\widetilde{SL}(2,\R)$ are isomorphic to $\Z$.)
However, a proof requires a different approach and might be given elsewhere.

\section{Non-global inextendible solutions, kinks}
\plabel{Kinks}

The solutions obtained so far have all been geodesically complete, or, if
not, the curvature or some physical field blew up at the boundary, rendering
a further extension impossible. However, these global spacetimes are not all
inextendible ones: it is, for instance, possible that the extremals are all
incomplete, the fields and the curvature scalar all remain finite, yet when
attempting to extend the solution one runs into problems, because the
extension would no longer be smooth or Hausdorff or similar
(cf.\ e.g.\ the Taub-NUT cylinders of Sec.\ \ref{Factor}).
The purpose of this section shall be to give some further
examples, to outline some general features of such solutions, and
to discuss to which extent a classification is possible.

A familiar example for the above scenario is the metric \cite{Dunn}
\begin{equation}
  g=e^{-2t}\left(-\cos2x\,dt^2-2\sin2x\,dt\,dx+\cos2x\,dx^2\right)\,,
  \plabel{flatkink}
\end{equation}
where the coordinate $x$ is supposed to be periodically wrapped up,
$x \sim x + n\pi$. These are $n$-kink solutions, which means that,
loosely speaking, the lightcone tilts upside-down $n$ times
when going along a non-contractible non-self\/intersecting loop on the
cylindrical spacetime. For $n=2$, however, \re{flatkink}) is nothing but
flat Minkowski space, the origin being removed, as is easily seen by
introducing polar coordinates
\begin{equation}
  \tilde x=e^{-t}\cos x\,,\quad \tilde t=e^{-t}\sin x
  \plabel{polar}
\end{equation}
into the metric
\begin{equation}
  g=d\tilde t^2-d\tilde x^2 \,,
\end{equation}
a fact that seems to have been missed in most of the literature.
Consequently, the metric is incomplete at the origin; it has a hole which
can easily be filled by inserting a point, leaving ordinary Minkowski space
without any kink.
For $n \ne 2$, on the other hand, (which are covering solutions of the above
punctured Minkowski plane, perhaps factored by a point-reflexion)
this insertion can no longer be done, because it would yield a `branching
point' (conical singularity) at which the extension could not be
smooth. Thus these manifolds are inextendible but nevertheless incomplete
and certainly the curvature does not diverge anywhere ($R\equiv0$).

Such a construction is of course possible for {\em any\/} spacetime,
leading to inextendible $n\ne2$-kinks.
However, if there is a Killing symmetry present, then even in the
2-kink situation inextendible metrics can be obtained.
To see this let us try to adapt the factorization approach of
the previous section to these kink-solutions. First of all, since a point
has been removed the manifold is
no longer simply connected, so one must pass to its universal covering,
which now winds around the removed point infinitely often in new layers
(cf.\ {\bf G3} versus {\bf G4} in Fig.\ 11).
All above kink-solutions can then be obtained by factoring out a `rotation'
of a multiple of $2\p$ (or $\p$, if there is a point-reflexion symmetry)
around the hole.
But according to the previous sections there should also
occur a kind of boost-parameter. Is it meaningful in this context? 

The answer to this question is yes, and this is perhaps best seen at the
flat 2-kink example, the Killing field chosen to describe boosts around the
(removed) origin. As long as the origin was supposed to belong to the manifold,
smoothness singled out one specific boost value for the gluing of the
overlapping sectors, leading to Minkowski space; otherwise there would have
occurred a conical singularity at the origin. However, if this point is
removed, then there is no longer any restriction on the boost-parameter. Its
geometrical meaning is that after surrounding the origin a boost has to be
applied before gluing or, in terms of fundamental regions, that a wedge has
to be removed from the original (punctured) Minkowski space and the
resulting edges are glued by the boost (also the tangents must be mapped
with the tangential map of this boost, cf.\ Fig.\ 21\ ({\it a\/})).
Of course, it is also possible to insert a wedge, but this is
equivalent to removing a wedge from an adjacent (stationary) sector.
\begin{figure}[htb]
\begin{center}
\leavevmode
\epsfxsize 12cm \epsfbox{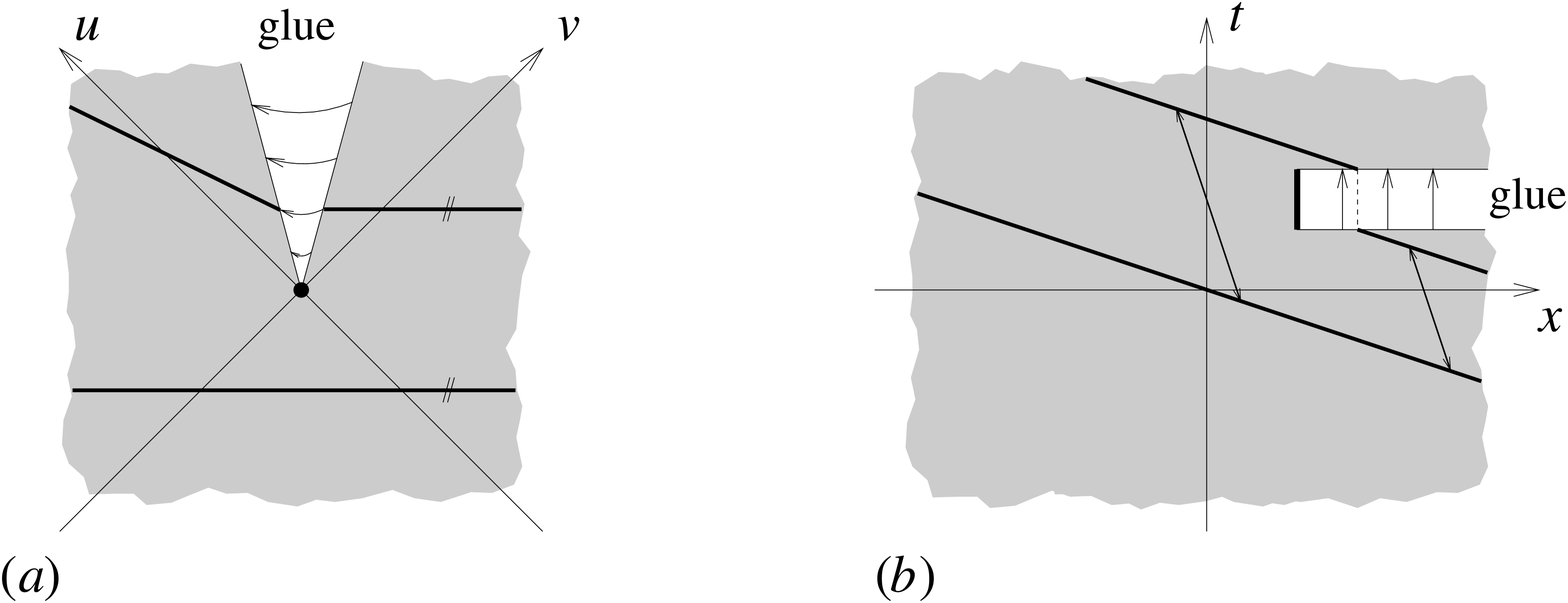}
\end{center}
\renewcommand{\baselinestretch}{.9}
\small \normalsize
\begin{quote}
{\bf Figure 21:} {\small ({\it a\/}) Minkowski kink with non-trivial
holonomy. This space can be obtained by removing a wedge from flat Minkowski
space and gluing together the corresponding boundary lines by a boost.
Due to this construction two extremals which are parallel on one side of the
origin are mutually boosted on the other side (cf.\ bold lines).
Thus the holonomy is non-trivial (surrounding the origin yields a boosted
frame), and at the origin there would occur a conical singularity. 
({\it b\/}) Another inextendible Minkowski kink; it has trivial holonomy but
the distance of parallels passing the hole changes.}
\end{quote}
\end{figure}
Clearly such a space is everywhere flat (except at the
origin, which is considered not to belong to the manifold) but has
non-trivial holonomy.
For instance, two timelike extremals which are parallel `before' passing
the origin at different sides will be mutually boosted afterwards
(bold lines in Fig.\ 21\ ({\it a\/})).

In the above example we chose as Killing symmetry the boosts centered at
the origin. However, Minkowski space also exhibits translation
symmetries. An analogous construction can be applied also in this
case with the following geometrical interpretation:
cut out a whole slit (in direction of the chosen translation), remove the
strip on one side of the slit, and glue together the resulting faces
(cf.\ Fig.\ 21\ ({\it b\/})). This manifold has now trivial holonomy,
but the metric distance of two generic parallels passing the hole
changes. Thus the manifold is so badly distorted that it cannot be completed
to ordinary Minkowski space, either.
In contrast to the former case this space can still be smoothly extended
further: one can simply continue beyond the remaining left edge of the slit
(bold line)
into an overlapping layer whose upper and lower faces have to be glued together
(since the endpoints of the slit have to be identified).
This yields a maximally extended cylinder, where the (identified) endpoints of
the slit constitute a conical singularity and should be removed.

It is relatively straightforward to write down the metric for the
above examples in a circular region around the hole (but not too close to the
hole), using smooth but non-analytic functions. % (like bump etc.)
Analytic charts are more difficult to obtain, but at least for the
case Fig.\ 21 ({\it b\/}) also this is possible
(cf.\ \cite{KINKpaper}). This last example, furthermore, can easily be
generalized to an arbitrary metric with
Killing symmetry: one has just to introduce Eddington-Finkelstein
coordinates \re{011h}) in one patch. Then the analogous construction
with $(x,t)\rightarrow(x^0,x^1)$ yields a one-parameter family of
inextendible 2-kink solutions (resp.\ $2n$-kink). Explicit charts can be
found in \cite{KINKpaper}.
There is (sometimes) even the possibility to introduce discrete
parameters like a `block-number': 
one could, e.g., take {\bf G9} and make a long horizontal slit
over a number of blocks, then remove a few blocks on the one side 
and glue together again (cf.\ Fig.\ 22).
\begin{figure}[htb]
\begin{center}
\leavevmode
\epsfxsize 10cm \epsfbox{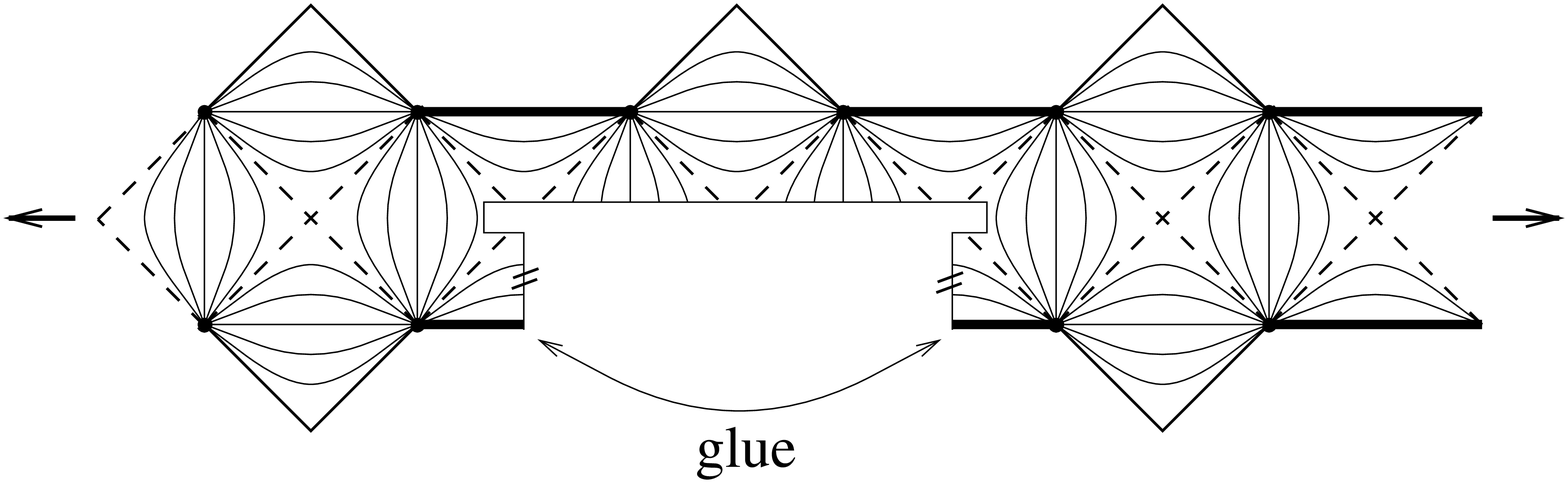}
\end{center}
\renewcommand{\baselinestretch}{.9}
\small \normalsize
\begin{quote}
{\bf Figure 22:} {\small Yet another kink for {\bf G8,9}.}
\end{quote}
\end{figure}  

\medskip

It is somewhat problematic to give a complete classification of the kink
solutions found above. Of course they could be described as factor spaces
of {\em limited\/} coverings of the original universal covering solution.
First, however, this would rather be a mere enumeration of the possible cases
than a classification. Secondly, we do not want to distinguish two solutions,
one of which is just an extension of the other. Thus we should only consider
maximally extended limited coverings. In the above examples they all
had only conical singularities, but it is not evident that this
should be the most general scenario.
Disregarding this question of the extension, the kinks are of
course characterized by their kink-number, a (real) boost-parameter, and
perhaps further discrete parameters (block-number or the like).  

Finally we want to mention that such surgery is not restricted to cylindrical
solutions (i.e.\ one hole only), but, within any of the global solution
obtained in the previous sections, one can cut any number of holes, each
giving rise to one boost-, one kink-, and perhaps some further discrete
parameter. And as is well-known from complex analysis (Riemann surfaces),
one can even obtain surfaces of higher genus (e.g.\ genus 1 with four
branching points, etc.) in that way.

\section{Conclusion}
\plabel{Concl}

We have succeeded in finding all global (as explained in Sec.\ \ref{Intro})
maximally extended solutions for
generalized dilaton gravity or, more general, for any gravity model
with Killing symmetry. The occurring topologies were found to
depend only on the number and (degeneracy-)degrees
of the Killing horizons within a corresponding Eddington-Finkelstein
coordinate patch \re{011h}). In particular, for three or more horizons
we obtained solutions on non-compact surfaces of arbitrary genus and
with an arbitrary nonzero number of holes.
Besides these global solutions, where a further extension was impossible
due to the
completeness of the extremals or divergent curvature or dilaton fields,
we have also found classes of incomplete inextendible solutions, where an
extension was impeded by a conical singularity or similar defects
(Sec.\ \ref{Kinks}).

As a general rule, the number of additional continuous parameters arising for
non-trivial topologies equals the rank of the fundamental group $\pi_1(M)$.
The dimension of the solution space (including the integration constant $C$)
exceeds this rank by one, certainly. In the more general case including
Yang-Mills fields, which has not been treated here explicitly, this dimension
is $(\mbox{rank }\pi_1(M)+1)(\mbox{rank (gauge group)}+1)$.

In this series the Poisson-$\s$-formalism was used only in the first
step, to obtain the local form of the metric \re{011h}). Then we proceeded
in a purely classical gravitational manner. It seems to be possible,
however, to stay all the way within the Poisson-$\s$-framework:
In this case the continuous parameter $C$ as well as the discrete labels
encountered above are reobtained as homotopy classes of maps of the
2D world-sheet (spacetime) into symplectic leaves in the target space.
The additional continuous (boost) parameters, on the other hand, are found
to correspond to generalized ``parallel transporters'' $\oint_\G A_{\wt1}$
around non-contractible loops $\G$ on the spacetime.
We hope to come back to this aspect in more detail elsewhere.

\section*{Acknowledgement}

We are grateful to H. Balasin, H.D. Conradi, S.R. Lau, F. Schramm,
and V. Schulz for discussions and to W. Kummer for his encouragement
during the long-lasting genesis of the present paper.
The work has been supported in part by the Austrian Fonds zur F\"orderung
der wissenschaftlichen Forschung (FWF), project P10221-PHY.


\vspace{1cm}

\begin{thebibliography}{00}
\bibitem{Odintsov} T.\ Banks and M.\ O'Loughlin, {\em Nucl. Phys.} 
  {\bf B362} (1991); 649, S.D.\ Odintsov and I.L.\ Shapiro, 
  {\em Phys.\ Lett.}    {\bf B263} (1991), 183. 
\bibitem{I} T.\ Kl\"osch and T.\ Strobl, {\em Class.\ Quantum Grav.}
  {\bf 13} (1996), 965; Corrigendum, {\em Class. Quantum Grav.} {\bf 14}
  (1997), 825. (Referred to as {\bf I} in the text).
\bibitem{Thm} Y.\ Choquet-Bruhat, C.\ DeWitt-Morette, {\em Analysis,
  Manifolds and Physics}, North-Holland Physics, 1982. 
\bibitem{II} T.\ Kl\"osch and T.\ Strobl, {\em Class.\ Quantum Grav.}
  {\bf 13} (1996), 2395. (Referred to as {\bf II} in the text).
\bibitem{Kat} M.O. Katanaev and I.V. Volovich,
  {\em Phys.\ Lett.} {\bf 175B} (1986), 413.
\bibitem{IV} T.\ Kl\"osch and T.\ Strobl, {\em Classical and Quantum
  Gravity in 1+1 Dimensions; Part}\nolinebreak\ {\bf IV}:
  {\em The Quantum Theory}, in preparation.
\bibitem{Briefetc}  
  T.\ Strobl, {\em Phys.\ Rev.} {\bf D50} (1994), 7346. \newline 
  P.\ Schaller and T.\ Strobl, \newline 
  {\em Quantization of Field Theories Generalizing Gravity-Yang-Mills
     Systems on the Cylinder}, in LNP {\bf 436},
     p.\ 98, `Integrable Models and Strings', eds.\ A. Alekseev et al,
     Springer 1994 or gr-qc/9406027. \newline
  {\em Mod.\ Phys.\ Letts.} {\bf A9} (1994), 3129. \newline
  {\em Poisson $\sigma$-models: A generalization of Gravity-Yang-Mills
     Systems in Two Dimensions}, in the Proceedings of the International
     Workshop on `Finite Dimensional Integrable Systems', p.\
     181-190, Eds.\ A.N.\ Sissakian and G.S.\ Pogosyan, Dubna 1995, 
     or   hep-th/9411163  \newline 
  {\em Introduction to Poisson $\s$-Models},  in LNP {\bf 469},  p.\ 321
     `Low-Dimensional Models in Statistical Physics and Quantum Field
     Theory', Eds.\ H.\ Grosse and L.\ Pittner, Springer 1996, or  
     hep-th/9507020.  
\bibitem{Ashbuch} A.\ Ashtekar, {\em Lectures on Non-Perturbative Canonical
  Gravity}, World Scientific, Singapore 1991.
\bibitem{PhysLetts} P.\ Schaller and T.\ Strobl, {\em Phys. Lett.} {\bf B337}
  (1994), 266.
\bibitem{Haw} S.W. Hawking and G.F.R. Ellis, {\em The Large Scale 
  Structure of Space-Time}, Cambridge University Press, 1973. 
\bibitem{Wolf} J.A.\ Wolf, {\em Spaces of Constant Curvature}, 
  McGraw-Hill, 1967.
\bibitem{Massey} W.S. Massey, {\em Algebraic Topology: An Introduction},
  GTM {\bf 56}, Springer, 1977.
\bibitem{CombGr} R.C. Lyndon, P.E. Schupp, {\em Combinatorial Group Theory},
  Springer, 1977.
\bibitem{Mis}  C.W. Misner in {\em Relativity Theory
  and Astrophysics I: Relativity and Cosmology}, ed.\ J.\ Ehlers,
  Lectures in Applied Mathematics, Vol.\ 8, p. 160, AMS 1967. 
\bibitem{Geroch} R.P. Geroch, {\em J. Math.\ Phys.} {\bf 9} (1968), 450.
\bibitem{Burckel} R.B. Burckel, {\em An Introduction to Classical Complex
  Analysis, Vol.\ I}, Birkh\"auser, 1979.
\bibitem{JT}  B.M. Barbashov, V.V. Nesterenko and
  A.M. Chervjakov, {\em Theor.\ Mat.\ Phys.} {\bf 40} (1979), 15;
  C. Teitelboim, {\em Phys.\ Lett.} {\bf B126} (1983), 41;
  R. Jackiw,  {\em 1984 Quantum Theory of Gravity}, ed S.\ Christensen
  (Bristol: Hilger), p.\ 403.
\bibitem{Dunn} K.A. Dunn, T.A. Harriott, J.G. Williams, {\em J. Math.\
  Phys.} {\bf 33} (1992), 1437;\\
  M. Vasilic, T. Vukasinac, {\em Class.\ Quantum Grav.} {\bf 13} (1996), 1995.
\bibitem{KINKpaper} T.\ Kl\"osch and T.\ Strobl, {\em A Global View of Kinks
  in 1+1 Gravity}, preprint, gr-qc/9707053.
\end{thebibliography}
\end{document}